\RequirePackage{fix-cm}
\documentclass[smallextended]{svjour3}       
\smartqed  
\usepackage{graphicx}
\journalname{J. Nonlinear Science}
\usepackage{amsmath,amssymb}
\usepackage{mathrsfs}
\usepackage{mathbbol}
\usepackage{mathtools}

\usepackage{changepage}

\usepackage[utf8x]{inputenc}

\usepackage{textcomp,marvosym}
\usepackage{cite}

\usepackage{caption}

\usepackage{nameref,hyperref}

\usepackage[right]{lineno}

\usepackage{microtype}
\DisableLigatures[f]{encoding = *, family = * }

\usepackage[table]{xcolor}

\usepackage{array}

\newcolumntype{+}{!{\vrule width 2pt}}

\usepackage{rotating}
\newcommand{\red}[1]{\textcolor[rgb]{1,0,0}{#1}}
\newcommand{\pink}[1]{\textcolor[rgb]{1,0,1}{#1}}
\newcommand{\blue}[1]{\textcolor[rgb]{0,0,0}{#1}}
\newcommand{\at}[2][]{#1|_{#2}}

\definecolor{pink}{HTML}{FFC0CB}
\definecolor{red}{HTML}{FF0000}

\newlength\savedwidth

\usepackage{graphicx}
\usepackage{scrextend}
\usepackage{url}
\usepackage{lipsum}
\usepackage[all]{xy}
\usepackage{xcolor}
\usepackage{authblk}

\usepackage{amsmath}
\usepackage{dsfont}
\usepackage{cancel}

\usepackage{pbox}
\usepackage{bbm}

\usepackage{blindtext}  
\addtokomafont{labelinglabel}{\sffamily}

\usepackage[normalem]{ulem}

\newcommand{\vect}[1]{\boldsymbol{#1}}

\usepackage[aboveskip=1pt,labelfont=bf,labelsep=period,justification=justified,singlelinecheck=off]{caption}

\renewcommand{\figurename}{Fig}

\date{}

\begin{document}

\title{Cortical Divisive Normalization from Wilson-Cowan Neural Dynamics
}


\author{Jes\'us Malo \and Jos\'e Juan Esteve-Taboada \and Marcelo Bertalm\'{i}o}


\institute{J. Malo \at
              Image Processing Lab, Universitat de Val\`{e}ncia, Spain \\
              Tel.: +34 963 544 099
              \email{jesus.malo@uv.es}           
           \and
           J.J. Esteve Taboada \at
              Image Processing Lab,
              Universitat de Val\`{e}ncia, Spain
           \and
           M. Bertalmío \at
              Spanish National Research Council (CSIC), Madrid, Spain
}
\vspace{-2cm}

\maketitle

\begin{abstract}
\vspace{-2.2cm}
Divisive Normalization and the Wilson-Cowan equations are well-known influential models of nonlinear neural interaction
[\emph{Carandini and Heeger} Nature Rev. Neurosci. 2012; \emph{Wilson and Cowan} Kybernetik 1973].
However, they have been always treated as different approaches, and have not been analytically related yet.

In this work we show that Divisive Normalization can be derived from the Wilson-Cowan dynamics.
Specifically, assuming that Divisive Normalization is the steady state of the Wilson-Cowan
differential equations, we find that the kernel that controls neural interactions in Divisive Normalization
depends on the Wilson-Cowan kernel but also depends on the signal.
A standard stability analysis of a Wilson-Cowan model with the parameters obtained
from our relation shows that the Divisive Normalization solution is a stable node.
This stability \blue{suggests the appropriateness} of our steady state assumption.

The proposed theory provides a mechanistic foundation 
for the suggestions that have been done on the need of signal-dependent Divisive Normalization
in [\emph{Coen-Cagli et al.} PLoS Comp.Biol. 2012].
Moreover, this theory explains the modifications that had to be introduced ad-hoc in Gaussian kernels of Divisive Normalization in [\emph{Martinez et al.} Front. Neurosci. 2019] to reproduce contrast responses in V1 cortex.
Finally, the derived relation implies that the Wilson-Cowan dynamics also reproduce visual masking and subjective image distortion,
which up to now had been explained mainly via Divisive Normalization.

\keywords{\small{Neural Networks \and Visual Cortex \and Nonlinear Neural Interactions \and Divisive Normalization \and Wilson-Cowan Equations \and Visual Masking \and Contrast Perception}}
\end{abstract}


\section{Introduction}

%

A number of perceptual experiences in different modalities can be described with the Divisive Normalization interaction among the outputs of linear neurons~\cite{Carandini94,Carandini12}. In particular, in vision, the perception of color~\cite{Brainard05}, texture~\cite{Watson97}, and motion~\cite{Simoncelli98} seem to be mediated by this nonlinear interaction. \blue{Intuitively, this \emph{divisive} interaction modifies the response of a sensor by normalizing it with the responses of the neighbor neurons, thus explaining inhibition by the surround.}

The discussion on the circuits underlying the Divisive Normalization in~\cite{Carandini12} suggests that there may be different architectures leading to this specific computation. Recent results suggest specific mechanisms for Divisive Normalization in certain situations~\cite{Carandini16}, but the debate on the physiological implementations is still open.
On the other hand, a number of functional advantages \cite{Schwartz09,Schwartz11,Coen12,Coen13} suggest that the kernel that describes the interaction in Divisive Normalization should be adaptive (i.e. signal or context dependent). Moreover, the match between the linear receptive fields and the interaction kernel in the Divisive Normalization is not trivial: the conventional Gaussian kernels in \cite{Watson97,Malo10} had to be tuned by hand to reproduce contrast responses \cite{Martinez19}.

These open questions imply that it is interesting to relate Divisive Normalization to other models of neural interaction for a better understanding of its implementation,
the structure of the interaction kernel, and its eventual dependence with the signal. Interesting possibilities to consider are the classical dynamic neural field models
of Wilson-Cowan \cite{Wilson72,Wilson73,Bressloff03}, Amari~\cite{Amari77}, or Grossberg~\cite{Grossberg68,Grossberg88}, \blue{ all of which have a similar subtractive nature~\cite{Chow20}: 
in subtractive models, the surround modifies the activity of a given sensor by substracting a weighted average of its neighbor's responses, as opposed to the division made in normalization models.}

Subtractive and divisive adaptation models have been qualitatively related before~\cite{Wilson93,Cowan02}.
Both models have been shown to have similar advantages in information-theoretic terms:
\blue{the Wilson-Cowan interaction in a neural layer uniformizes the  
probability density of the responses~\cite{Bertalmio14}, 
and there is less redundancy among them~\cite{Gomez19}. Similarly, Divisive Normalization layers reduce the relations 
between the responses~\cite{Malo06a}, factorize the joint probability of the responses~\cite{Malo10}, and maximize 
the transmitted information from the input~\cite{Malo20,Malo22}.}
Additionally, both models provide similar descriptions of pattern discrimination \cite{Wilson93,Bertalmio17}.
However, despite all these similarities, no direct analytical correspondence has been established between
these models yet.

In this paper, we assume that the psychophysical behavior described by Divisive Normalization comes from
underlying neural interactions that follow the Wilson-Cowan equations. In particular, we identify the Divisive Normalization response with the stationary regime of a Wilson-Cowan system.
From this identification we derive an expression for the Divisive Normalization kernel
in terms of the interaction kernel of the Wilson-Cowan equations.

This analytically derived relation has the following interesting consequences:
\vspace{0.2cm}

\blue{(1) It has been suggested that Divisive Normalization should depend on the input because of functional reasons~\cite{Coen12,Coen13}, but no physiological mechanism was proposed to implement this statistical adaptation. 
The proposed relation of the Divisive Normalization with a dynamical system with fixed wiring among neurons provides a mechanistic explanation for this dependence with the input.} 

%
%
%

(2) The relation explains the modifications that had to be introduced ad-hoc in the kernel of Divisive Normalization in \cite{Martinez19} to reproduce contrast responses. This implies that the Wilson-Cowan dynamics reproduce visual
masking, which up to now had been mainly explained via Divisive Normalization \cite{Foley94,Watson97}.

(3) The relation allows to build effective image quality metrics based on the Wilson-Cowan model, something which, to the best of our knowledge, hasn't been considered before in the literature, as opposed to the many examples of metrics based on Divisive Normalization~\cite{Teo94,Laparra10a,Malo10,Laparra17b,Hepburn2020}.

(4) A standard stability analysis of a Wilson-Cowan model with the parameters obtained from our relation shows that the Divisive Normalization solution is a stable node of the dynamic model.
This shows the \blue{appropriateness} of our steady state assumption.
Moreover, this stability justifies the straightforward use of Divisive Normalization with time-varying stimuli, as in~\cite{Simoncelli98}.

\vspace{0.3cm}
The structure of the paper is as follows. 
The \emph{Materials and Methods} section reviews the retina-V1 neural path and the contrast perception of visual patterns. We also introduce the notation of the models: the Divisive Normalization and the Wilson-Cowan equations. Besides, we recall some experimental facts that will be used to illustrate the performance of the proposed relation: (1)~contrast responses curves imply certain interactions between subbands~\cite{Cavanaugh00,Watson97}, (2)~the Divisive Normalization kernel must have a specific structure (identified in~\cite{Martinez19}) to reproduce contrast response curves, and (3)~the shape of the Divisive Normalization kernel should have a specific dependence with the surrounding signal~\cite{Cavanaugh02a, Cavanaugh02b}.
In the \blue{\emph{Analytical Results}} section we derive the relation between the Divisive Normalization and the Wilson-Cowan equations based on the steady state assumption.
The \blue{\emph{Numerical Experiments}} section illustrates with simulations the mathematical properties and the perceptual consequences of the proposed relation.
First, \blue{we experimentally check the convergence} of the Wilson-Cowan solution to the Divisive Normalization response \blue{in a wide range of model parameterizations.} 
\blue{We quantify the error introduced by the approximations done to get the analytical results.}
Moreover, we illustrate the appropriateness of the steady state assumption by showing that the Divisive Normalization is a stable node of the Wilson-Cowan system. Then, we address contrast perception facts using the proposed relation to build a psychophysically meaningful Wilson-Cowan model: we theoretically derive the specific structure of the kernel that was previously inferred empirically \cite{Martinez19},
we show that the proposed interaction kernel adapts with the signal, and as a result, we reproduce general trends of contrast response curves. Finally, we discuss the use of the derived kernel in predicting the perceptual metric of the image space.
The \emph{Final Remarks} section concludes the paper.

%
%
%
%

\section{Materials and Methods}

In this work the theory is illustrated in the context of models of the retina-cortex pathway.
The considered framework follows the approach suggested in~\cite{Carandini12}:
a cascade of four isomorphic \emph{linear+nonlinear} modules.
These four modules sequentially address brightness, contrast, frequency filtered contrast masked in the spatial domain,
and orientation/scale masking. An example of the transforms of the input in such models is shown in Fig.~1.

In this general context we focus on the cortical (fourth) layer: a set of linear sensors with wavelet-like receptive fields
modelling simple cells in V1, and the nonlinear interaction between the responses of these linear sensors.
Divisive Normalization has been the conventional model used for the nonlinearity to describe contrast perception
psychophysics \cite{Watson97}, but here we will explore the application of the Wilson-Cowan model in the contrast perception context.

Below we introduce the notation of both models of neural interaction and the facts on contrast perception that should be explained by the models.

\begin{figure}[!t]
	\centering
    \small
    \setlength{\tabcolsep}{2pt}
    \vspace{-0cm}
    \begin{tabular}{c}
    \hspace{-0.0cm}  \includegraphics[height=0.78\textheight]{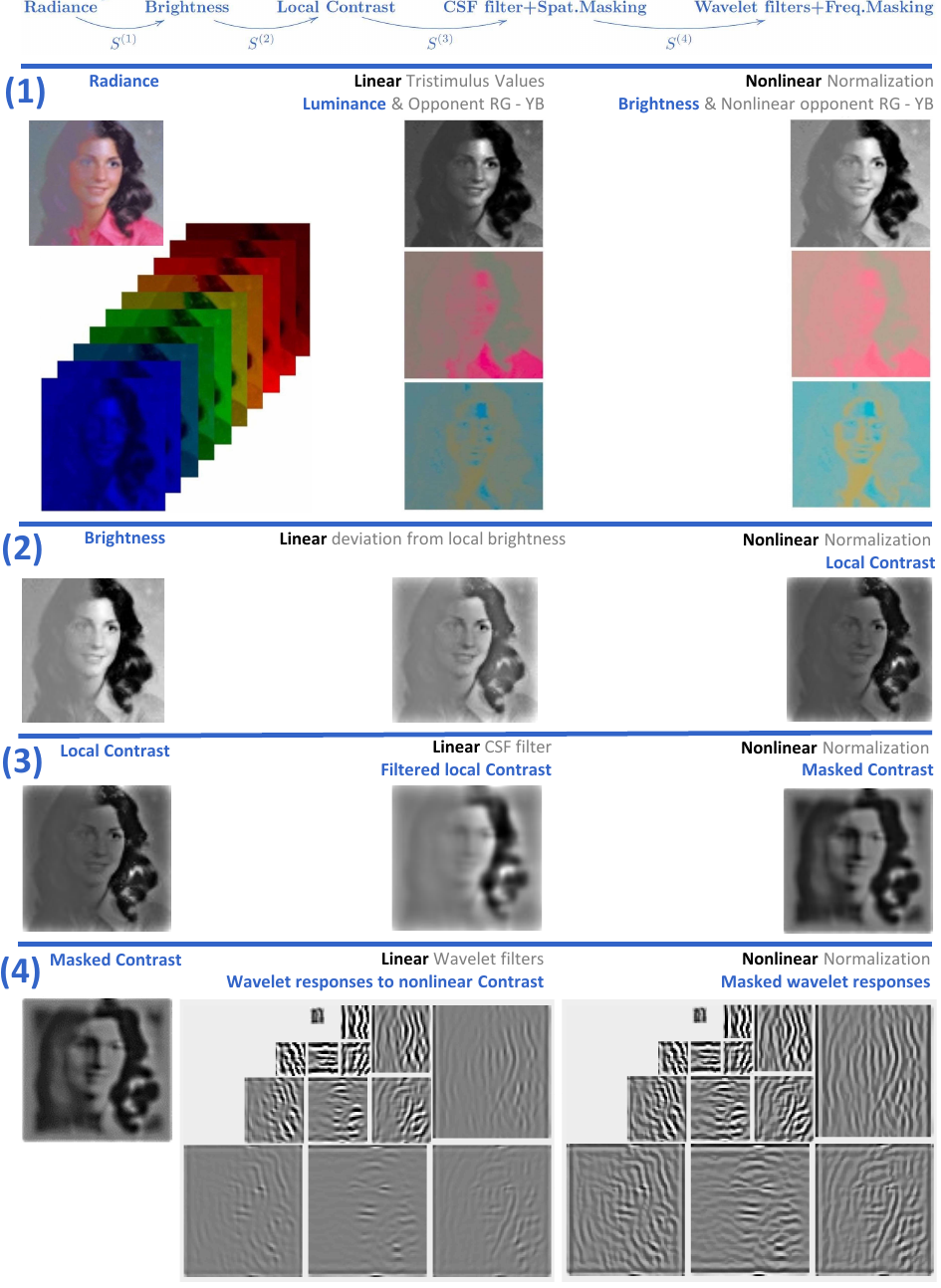} \\
    \vspace{-0.0cm}
    \hspace{-0.0cm}\pbox{1.4\textwidth}{\textbf{Fig 1.} \scriptsize{\textbf{Signal transforms in the retina-cortex pathway: a cascade of linear+nonlinear modules} (example from~\cite{Martinez17}).}
                         The input is the spatial distribution
                         of the \emph{spectral irradiance} at the retina.
                         (1)~The linear part of the first layer consist of three positive spectral sensitivities (Long, Medium, Short, LMS, wavelengths) and a linear recombination of the LMS values with positive/negative weights. This leads to three tristimulus values in each spatial location: one of them is proportional to the luminance, and the other two have opponent chromatic meaning (red-green and yellow-blue). These linear tristimulus responses undergo adaptive saturation transforms.
                         Perception of \emph{brightness} is mediated by an adaptive Weber-like nonlinearity applied to the luminance at each location. This nonlinearity enhances the response in the regions with small linear input (low luminance).
                         (2)~The linear part of the second layer computes the deviation of the brightness at each location from the local brightness. Then, this deviation is nonlinearly normalized by the local brightness
                         to give the local contrast.
                         (3)~The responses to local contrast are convolved by center surround receptive fields (or filtered by the Contrast Sensitivity Function). Then the linearly filtered contrast is nonlinearly normalized by the local contrast. Again normalization increases the response in the regions with small input (low contrast).
                         (4)~After linear wavelet transform, each response is nonlinearly normalized by the activity of the neurons in the surround. Again, the activity relatively increases in the regions with low input. The common effect of the nonlinear modules throughout the network is response equalization.}
                         \end{tabular}
    \vspace{-0.5cm}
\end{figure}
\setcounter{figure}{1}

\subsection{Modelling cortical interactions}


\blue{Our focus here is the last \emph{linear+nonlinear} module of the retina-V1 cascade in Fig.~1, and specifically the nonlinear layer that describes the interactions in the primary visual cortex V1.
The \emph{driving input} of this final nonlinear layer is the vector of energies, $\vect{e}$, of the responses of linear wavelet-like simple cells, and the \emph{output} of this interaction is the vector of nonlinear responses $\vect{x}$:}
\vspace{-0.2cm}
\begin{equation}
  \blue{\xymatrixcolsep{2pc}
  \xymatrix{ \vect{e}  \,\,\,\, \ar@/^0.7pc/[r]^{\scalebox{0.85}{$\mathcal{N}$}} & \,\,\,\, \vect{x}
  }}
  \label{global_response}
\end{equation}
\blue{In this work the two models considered describe the interaction $\mathcal{N}$ between the linear simple cells in V1. The Wilson-Cowan equations model neural firing rate dynamics that may converge to a
steady state. If that is the case, the long-term behavior of the Wilson-Cowan
equations may be similar to the Divisive Normalization model, since the latter models static neural firing rates.}



\vspace{-0.0cm}
\subsection{The Divisive Normalization model}
\paragraph*{Forward transform.}  \hspace{0.15cm} The conventional expression of the canonical Divisive Normalization~\cite{Carandini12} uses an element-wise formulation: 
\begin{equation}
        \blue{x_i = k_i \,\, \frac{e_i}{b_i + \sum_j H_{ij} e_j}}
        \label{DN_B0}     
\end{equation}
where the output vector of nonlinear activations, $\vect{x}$, depends on the energy of the input linear wavelet responses, $\vect{e}$, which are dimension-wise normalized by a sum of neighbor energies of the input.
For convenience for the derivations below, the transform can be rewritten in matrix form~\cite{Martinez17,Martinez19}:
\begin{equation}
    \vect{x} = \mathbb{D}_{\vect{k}} \cdot \mathbb{D}^{-1}_{\left( \vect{b} + \vect{H} \cdot \vect{e} \right)} \cdot \vect{e}
    \label{DN_B}
\end{equation}
\vspace{0.1cm}
where $\mathbb{D}_{\vect{v}}$ are diagonal matrices with the vector $\vect{v}$ in the diagonal. The non-diagonal nature of the interaction kernel $\vect{H}$ which is in the denominator, $\vect{b} + \vect{H} \cdot \vect{e}$, implies that the $i$-th element of the response is attenuated by the activity of the neighbor sensors, $e_j$ with $j\neq i$.
Each row of the kernel $\vect{H}$ describes how the energies of the neighbor simple cells attenuate each simple cell after the interaction.
Each element of the vectors $\vect{b}$ and $\vect{k}$ represents the semisaturation and the dynamic range of the nonlinear response of each sensor, respectively.
This nonlinear interaction only affects the amplitude of the responses, not its sign. As a result, for simplicity in the notation, throughout the work $\vect{x}$ refers to the vector of absolute values of the responses. The sign of the normalized responses is inherited from the sign of the linear wavelet responses.

\vspace{-0.0cm}
\paragraph*{Inverse transform.}  \hspace{0.15cm} The matrix notation~\cite{Martinez17,Martinez19} is convenient to derive the analytical inverse of the Divisive Normalization,
which will be used to obtain the relation between the two models considered in this work. The inverse is given by \cite{Malo06a,Martinez17,Martinez19}:
\begin{equation}
      \vect{e} = \left( I - \mathbb{D}^{-1}_{\vect{k}}\cdot\mathbb{D}_{\vect{x}}\cdot \vect{H} \right)^{-1} \cdot \mathbb{D}_{\vect{b}} \cdot \mathbb{D}^{-1}_{\vect{k}} \cdot \vect{x}
      \label{invDN}
\end{equation}
This inverse, originally proposed in the context of image coding~\cite{Malo06a}, has been used in other applications that require the reconstruction of the image~\cite{Camps08,Laparra17b}.

\vspace{-0.0cm}
\subsection{The Wilson-Cowan model}
The Wilson-Cowan dynamical system was proposed \blue{as a general model of the inhibitory/excitatory interactions between neural populations, and as application, it can be used to model the signal at specific stages in the visual pathway \cite{Wilson72,Wilson73,Bressloff03}}.
In Wilson-Cowan models, part of the neural population (part of the coefficients in the vectors $\vect{e}$ and $\vect{x}$) is \emph{excitatory} and part is \emph{inhibitory}, meaning how their magnitude affects the neighbors (in additive or subtractive way respectively).
Excitatory and inhibitory populations will be referred to as $\vect{e^e}$, $\vect{x^e}$, and $\vect{e^i}$, $\vect{x^i}$, respectively. We will consider that these excitatory and inhibitory neurons (or coefficients)
are interleaved in the vectors that describe the responses. Or, for simplicity, one may also represent them as separate rows in the response vectors:
\vspace{-0.25cm}
\begin{equation}
  \xymatrixcolsep{2pc}
   \begin{pmatrix} \vect{e^e} \\\vect{e^i} \end{pmatrix}
   \xymatrix{\,\,\, \ar@/^0.7pc/[r]^{\scalebox{0.85}{$\mathcal{N}$}} & \,\,\,
  }
  \begin{pmatrix} \vect{x^e} \\\vect{x^i} \end{pmatrix}
  \label{global_response2}
\end{equation}
In any case, the arbitrary arrangement of the neurons in the vectors does not
restrict the generality of the formulation. The only effect of this choice is the interpretation of the elements of the matrices that will represent the interaction between the neurons.

\paragraph*{Dynamical system.}  \hspace{0.15cm}  In Wilson-Cowan models~\cite{Wilson72,Wilson73,Bressloff03} the transform $\mathcal{N}$ is defined by differential equations that describe the temporal variation of the activity of the populations. In particular, this variation is driven by three factors:
\begin{itemize}
      \item An external driving input (either $\vect{e^e}$ or $\vect{e^i}$), in our case the responses of the linear V1 cells.
      \item The variation of the response of a population is auto-attenuated due to its own activity.
      \item The variation of the response is amplified by the excitatory responses and is moderated by the inhibitory responses.
\end{itemize}

Specifically, if in the notation of~\cite{Bressloff02a,Bressloff02b,Bressloff03}, \blue{which considers no refractory period in V1 neurons}, we explicitly identify the excitatory and inhibitory populations as done originally in~\cite{Wilson73},
for a neuron tuned to the feature $p$, we have one of these (excitatory or inhibitory) equations:
\begin{eqnarray}
      \hspace*{-1cm}\frac{\partial \, x^e_p(t)}{\partial t} &=& e^e_p(t) - \alpha^e_p \, x^e_p(t) + \int W^{ee}_{pp'} \, f(x^e_{p'}(t)) \, dp' - \int W^{ei}_{pp'} \, f(x^i_{p'}(t)) \, dp' \nonumber \\
      \hspace*{-1cm}\frac{\partial \, x^i_p(t)}{\partial t} &=& e^i_p(t) - \alpha^i_p \, x^i_p(t) + \int W^{ie}_{pp'} \, f(x^e_{p'}(t)) \, dp' - \int W^{ii}_{pp'} \, f(x^i_{p'}(t)) \, dp'
      \label{EqWCclassic}
\end{eqnarray}
or, in matrix notation:
\begin{eqnarray}
      \vect{\dot{x^e}} &=& \vect{e^e} - \mathbb{D}_{\vect{\alpha^e}} \cdot \vect{x^e} + \vect{W^{ee}} \cdot f(\vect{x^e}) - \vect{W^{ei}} \cdot f(\vect{x^i}) \nonumber \\
      \vect{\dot{x^i}} &=& \vect{e^i} - \mathbb{D}_{\vect{\alpha}^i} \cdot \vect{x^i} + \vect{W^{ie}} \cdot f(\vect{x^e}) - \vect{W^{ii}} \cdot f(\vect{x^i})
      \label{EqWC}
\end{eqnarray}
where $\vect{W^{ee}}$, $\vect{W^{ei}}$, $\vect{W^{ie}}$, $\vect{W^{ii}}$ are the matrices that describe
the excitatory and inhibitory relations between sensors, the activation function $f(\cdot)$ is a dimension-wise saturating nonlinearity, and the elements of the vectors $\vect{\alpha^e}$ and $\vect{\alpha^i}$ are the auto-attenuation parameters.
The above matrices are usually considered to be a fixed set of connections (wired relations), made of positive and negative Gaussian neighborhoods, that represent the local interaction between sensors \cite{Bressloff02b,Bressloff03,Faugueras09}.
Also note that, if in Eqs.~\ref{EqWC}, the inhibitory and the excitatory components are stacked together into a single vector (with some sort of arrangement as in Eq.~\ref{global_response2}), the two equations in the traditional Wilson-Cowan formulation can be represented by a single expression, as in~\cite{Bressloff03,Chow20}, here in matrix form:
\begin{eqnarray}
      \vect{\dot{x}} &=& \vect{e} - \mathbb{D}_{\vect{\alpha}} \cdot \vect{x} - \vect{W} \cdot f(\vect{x})
      \label{EqWC2}
\end{eqnarray}
where,
\vspace{-0.5cm}
\begin{equation}
      \vect{\alpha} = \begin{pmatrix} \vect{\alpha^e}\\\vect{\alpha^i} \end{pmatrix} \,\,\,\,\,\, , \,\,\,\,\,\,\,\,\,\,\,
      f(\vect{x}) = f\begin{pmatrix} \vect{x^e}\\ \vect{x^i} \end{pmatrix} 
      \,\,\,\,\,\, , \,\,\,\,\,\,\,\,\,\,\,
      \vect{W} = \begin{pmatrix} -\vect{W^{ee}} & \vect{W^{ei}} \\ -\vect{W^{ie}} & \vect{W^{ii}}   \end{pmatrix} \nonumber
\end{equation}
The above single-equation matrix formulation of the Wilson-Cowan model, Eq.~\ref{EqWC2}, is convenient to get the relation between the models, \blue{and clearly shows the subtractive nature of the interactions in the kernel $\vect{W}$ as opposed to 
the divisive nature of the interactions due to the kernel $\vect{H}$ in Eq.~\ref{DN_B}}.


\paragraph*{Steady state and inverse.} \hspace{0.15cm} The stationary solution of the above differential equation (obtained by taking $\vect{\dot{x}} =0$ in Eq.~\ref{EqWC2}) leads to the following analytical inverse for \blue{static} inputs:
\begin{equation}
      \vect{e} = \mathbb{D}_{\vect{\alpha}} \cdot \vect{x} + \vect{W} \cdot f(\vect{x})
      \label{invWC}
\end{equation}

As we will see in the \emph{Analytical Results} section, the identification of the decoding equations corresponding to both models, Eq.~\ref{invDN} and Eq.~\ref{invWC}, is the key to obtain
simple relations between their corresponding parameters.
%

\subsection{Experimental facts}

\subsubsection{Adaptive contrast response curves}
\begin{figure}[b]
	\centering
    \small
    \setlength{\tabcolsep}{2pt}
    \begin{tabular}{c}
    \hspace{0.5cm} \includegraphics[width=0.95\textwidth]{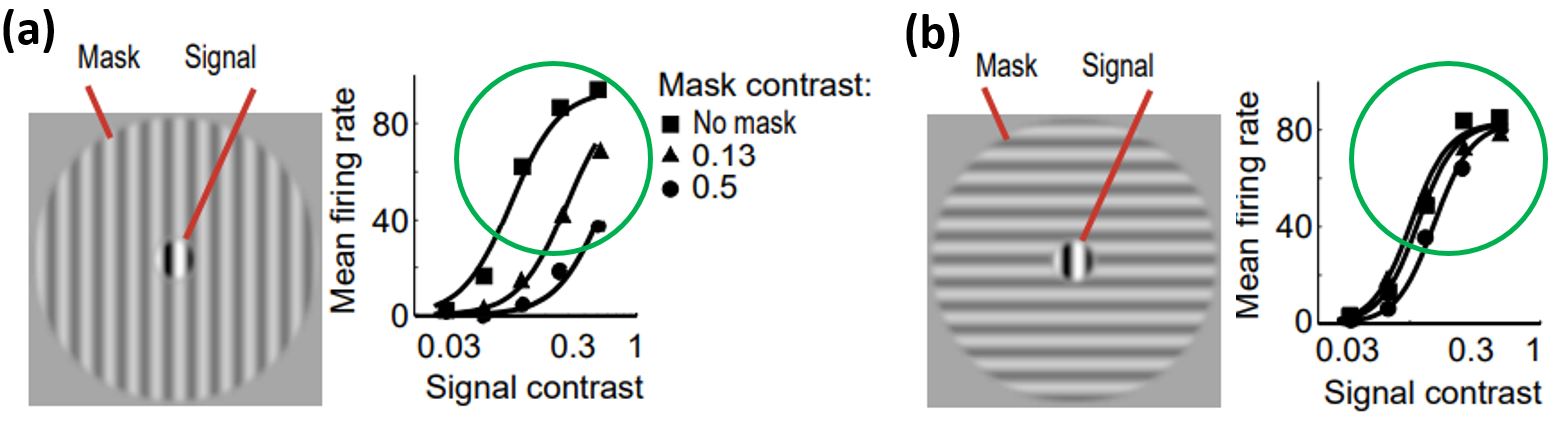} \\
    \end{tabular}
    \vspace{-0.0cm}
	\caption{\small{\textbf{Adaptive contrast response curves}.
    Mean firing rate (response) of V1 neurons tuned to the signal as a function of the signal contrast in two masking
    situations (adapted from \cite{Schwartz01,Cavanaugh00}).
    Note the decay in the response when signal and mask do have the same
    spatio-frequency characteristics (a), as opposed to the case where they do not (b). For visualization, the differences in the curves are highlighted by the green circles}.
}\label{facts}
    \vspace{-0.15cm}
\end{figure}

In the considered spatial vision context, the models should reproduce
the fundamental trends of contrast perception.
Thus, the slope of the contrast response curves should depend on the spatial frequency, so that
the sensitivity at threshold contrast is different for different spatial frequencies according to the Contrast Sensitivity Function (CSF) \cite{Campbell68}.
Also, the response curves should saturate with contrast \cite{Legge80,Legge81}.
Finally, the responses should attenuate with the energy of the background or surround, and this additional saturation should depend on the texture of the background \cite{Foley94,Watson97}:
if the frequency/orientation of the test is similar to the frequency/orientation of the background,
the decay should be stronger.
This background-dependent adaptive saturation, or \emph{masking}, is mediated by cortical sensors tuned to spatial frequency with responses that saturate depending on the background, as
illustrated in Fig. \ref{facts}.

The above trends are key to discard too simple models, and also to propose the appropriate modifications
in the model architecture to get reasonable results \cite{Martinez19}.

\subsubsection{Unexplained kernel structure in Divisive Normalization}

In the Divisive Normalization setting, the masking interaction between tests and backgrounds
of different textures is classically described by using a Gaussian kernel
in the denominator of Eq.~\ref{DN_B} in wavelet-like domains:
the effect of the $j$-th wavelet sensor on the attenuation of the $i$-th wavelet sensor
decays with the distance in space between the $i$-th and $j$-th sensors, but also with the spatial frequency and orientation \cite{Watson97}.
We will refer to these unit-norm Gaussian kernels as Watson and Solomon kernels \cite{Watson97}, and will be represented by $\vect{H}^{\vect{ws}}$.
Gaussian kernels are useful to describe the general behavior shown in Fig. \ref{facts}:
activity in close neighbors lead to strong decays in the response, while activity in neighbors
tuned to more distant features has smaller effect.

However, in order to have well behaved responses in every subband with every possible
background, a \emph{special balance} between the wavelet representation and the Gaussian kernels
is required.
When using reasonable log-polar Gabor basis or steerable filters to model V1 receptive fields, as in \cite{Watson97,Schwartz01}, the energies of the sensors tuned to low frequencies is notably higher than the energy of high-frequency sensors.
Moreover, the smaller number of sensors in low frequency subbands in this kind of wavelet representations
implies that unit-norm Gaussian kernels have bigger values in coarse subbands.
These two facts overemphasize the impact of low-frequency responses on high-frequency responses.
Thus, in \cite{Martinez19} we found that classical unit-norm Gaussian kernels require \emph{ad-hoc} extra modulation
to avoid excessive effect of low frequency backgrounds on high frequency tests.
The appropriate wavelet-kernel balance was then reestablished by introducing
extra high-pass filters in the Gaussian kernel $\vect{H}^{\vect{ws}}$, with the aim to moderate the effect of low frequencies \cite{Martinez19}:
\begin{equation}
      \vect{H} = \mathbb{D}_{\vect{l}} \cdot \vect{H}^{\vect{ws}} \cdot \mathbb{D}_{\vect{r}}
      \label{new_kernel_eq}
\end{equation}

In this new definition of the kernel: (1) the diagonal matrix at the right, $\mathbb{D}_{\vect{r}}$, pre-weights the subbands of $\vect{e}$ to moderate the effect of low frequencies
before computing the interaction; and (2) the diagonal matrix at the left, $\mathbb{D}_{\vect{l}}$, sets the relative weight of the masking for each sensor,
moderating low frequencies again.
The vectors $\vect{r}$ and $\vect{l}$ were tuned \emph{ad-hoc} in \cite{Martinez19} to get reasonable contrast response curves, both
for low and high frequency tests.


\vspace{0.2cm}
However, \emph{what is the explanation for this specific structure of the kernel matrix in Eq.~\ref{new_kernel_eq}?} \emph{And where do these two high-pass diagonal matrices come from?}

\subsubsection{Adaptive nature of kernel in Divisive Normalization}

\begin{figure}[!b]
	\centering
    \small
    \setlength{\tabcolsep}{2pt}
    \begin{tabular}{c}
    \hspace{-0cm} \includegraphics[width=1.1\textwidth]{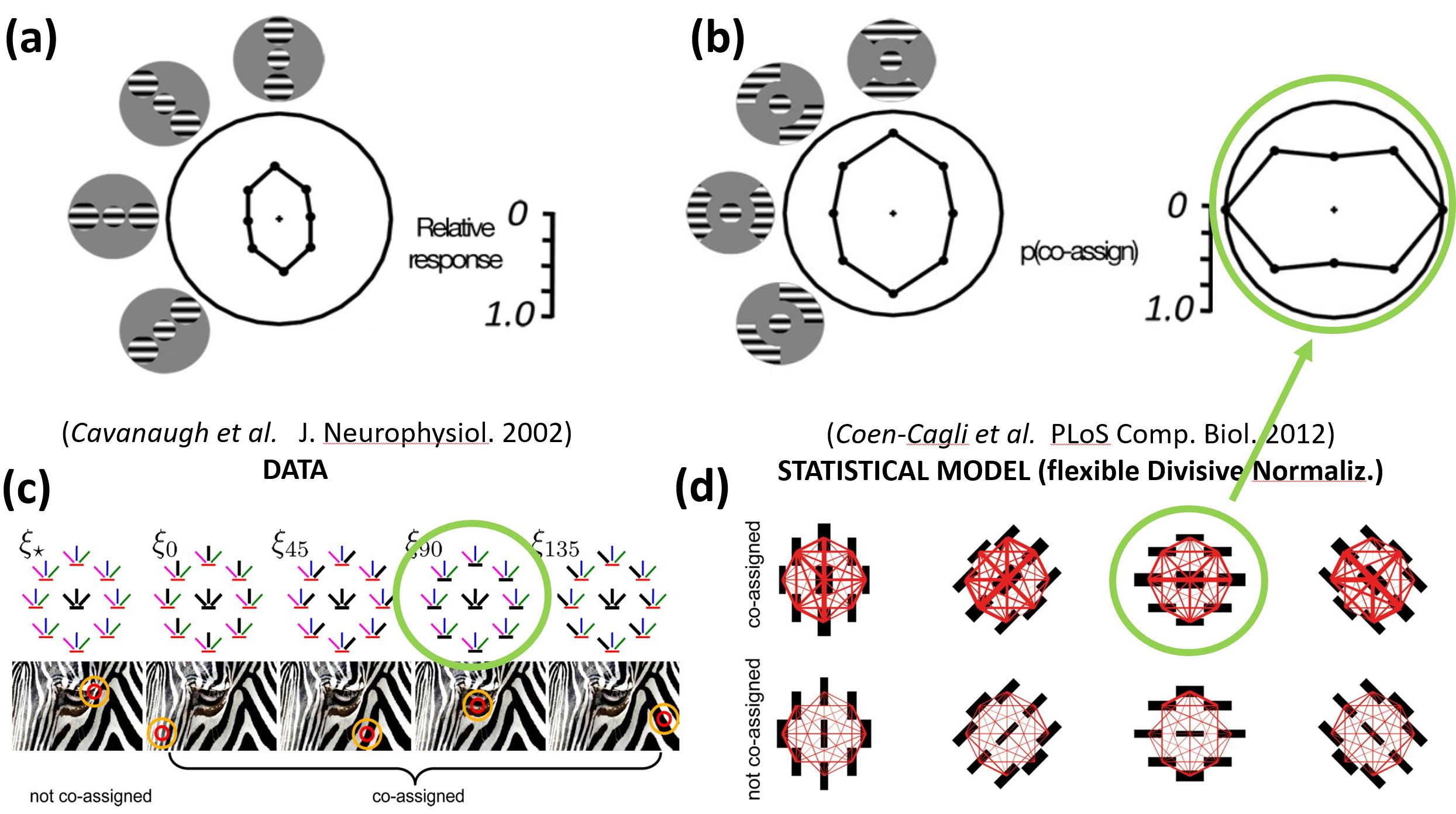} \\
    \end{tabular}
    \vspace{-0.0cm}
	\caption{\small{\textbf{Experimental context-dependent interaction \cite{Cavanaugh02b} and statistical model~\cite{Coen12}.} \textbf{(a)} Results of Cavanaugh et al. \cite{Cavanaugh02b}: images with a gray background represent the stimuli. Cell relative responses are shown as points inside the black normalization circle. The distance from the origin indicates the magnitude of the response, while its angle represents the location of the surrounding stimulus.
	\textbf{(b)} Cell response predicted from the statistical model of Coen-Cagli et al. \cite{Coen12},
	and probability that center and surround receptive fields are \emph{co-assigned} to the same normalization pool and contribute to the divisive normalization of the model response. The probability of co-assignment depends on the covariance with the surround, as shown below.
	\textbf{(c)} Different center-surround visual neighborhoods in a natural scene. In each case, the activity of the sensors in the surround
can be co-assigned to the activity in the center (i.e. considered in the normalization pool) if the orientation of maximally responding sensors is consistent (which is the case for four of the considered regions and not the case for the first).
The horizontal surround that is \emph{co-assigned} with the corresponding center is highlighted in bold. \textbf{(d)} Covariance matrices learned from natural images determine co-assignment: the orientation and relative position of the receptive fields are represented by the black bars (the thickness of the bar is proportional to the variance, while the thickness of the red lines is proportional to the covariance between the two connected bars).
\blue{This figure has been adapted from~\cite{Coen12}}.}}
	\label{fig_cohen}
    \vspace{-0.15cm}
\end{figure}

Previous physiological experiments on cats and macaques demonstrated that the effect of the surround on each cell does not come equally from all peripheral regions, showing up the existence of a spatially asymmetric surround \cite{Nelson1985, Deangelis94, Walker99, Cavanaugh02a, Cavanaugh02b}.
As shown in Fig. \ref{fig_cohen}.a, the experimental cell response is suppressed due to the surround, and this attenuation is greater when the grating patches are \emph{iso-oriented} and at the ends of the receptive field (as defined by the axis of preferred orientation) \cite{Cavanaugh02b}.

In the Divisive Normalization context, this asymmetry could be explained with non-isotropic interaction kernels. Depending on the texture of the surround, the interaction strength in certain direction may change. This would change the denominator, and hence the gain in the response.

Coen-Cagli et al. \cite{Coen12} proposed a specific statistical model to account for these contextual dependencies. This model includes grouping and segmentation of neighboring oriented features, and leads to a flexible generalization of the Divisive Normalization. Representative center-surround configurations considered in the statistical model are shown in Fig. \ref{fig_cohen}.c. A surround orientation can be either \emph{co-assigned} with the center group or \emph{not co-assigned}.
In the first case, the model assumes dependence between center and surround, and includes them both in the normalization pool for the center. In the second case, the model assumes center-surround independence, and does not include the surround in the normalization pool. Fig. \ref{fig_cohen}.d shows the covariance matrices learned from natural images between the variables associated with center and surround in the proposed statistical model. As expected, the variances of the center and its \emph{co-linear} neighbors, and also the covariance between them, are larger, due to the predominance of \emph{co-linear} structures in natural images. The cell response that is computationally obtained assuming their statistical model is shown in Fig. \ref{fig_cohen}.b, together with the probability that center and surround receptive fields are \emph{co-assigned} to the same normalization pool, and contribute then to the divisive normalization of the model response. Note that the higher the probability of \emph{co-assignment} between the center and surround, the higher the suppression (or the lower the signal) in the cell response.

This flexible (or adaptive) Divisive Normalization model based on image statistics \cite{Coen12} allows to explain the experimental asymmetry in the center-surround modulation \cite{Cavanaugh02b}. However, no direct mechanistic approach has been proposed yet to describe how this adaptation in the Divisive Normalization kernel may be implemented.

\section{Analytical Results: \blue{relation} between models}

The kernels that describe the relation between sensors in the Divisive Normalization and the Wilson-Cowan models, $\vect{H}$ and $\vect{W}$, have similar qualitative roles: both moderate the response, either by division or subtraction, taking into account the activity of the neighbor sensors.

In this section, we derive the  \blue{relation} between both models assuming that the Divisive Normalization behavior corresponds to the steady state solution of the Wilson-Cowan dynamics. This leads to an interesting analytical relation between both kernels, $\vect{H}$ and $\vect{W}$.

Under the steady state assumption, it is possible to identify the different terms in the decoding equations in both cases (Eq.~\ref{invDN} and Eq.~\ref{invWC}).
However, just to get a simpler analytical relation between both kernels, we make one extra simplification on each model.
%
\blue{Numerical experiments in the next section with natural inputs and a wide range of model parameterizations show that these simplifications are acceptable in practice.}

First, in the Divisive Normalization model, Eq.~\ref{invDN}, the identification may be simpler by taking the series expansion of the inverse.
This expansion was used in \cite{Malo06a} because it clarifies the condition for invertibility:
\begin{equation}
      \left( I - \mathbb{D}^{-1}_{\vect{k}} \cdot \mathbb{D}_{\vect{x}} \cdot \vect{H} \right)^{-1}
      = I + \sum_{n=1}^{\infty} \left( \mathbb{D}^{-1}_{\vect{k}} \cdot \mathbb{D}_{\vect{x}} \cdot \vect{H} \right)^n
      \nonumber
\end{equation}

The inverse exist if the eigenvalues of $\mathbb{D}^{-1}_{\vect{k}} \cdot \mathbb{D}_{\vect{x}} \cdot \vect{H}$ are smaller than one so that the series converges. In fact, if the eigenvalues are small, the inverse can be well approximated by a small number of terms in the Maclaurin series. Taking into account this approximation, Eq. \ref{invDN} may be written as:

\begin{eqnarray}
      \vect{e} & = & \mathbb{D}_{\vect{b}} \cdot \mathbb{D}^{-1}_{\vect{k}} \cdot \vect{x} + \left( \mathbb{D}^{-1}_{\vect{k}} \cdot \mathbb{D}_{\vect{x}} \cdot \vect{H} \right) \cdot \mathbb{D}_{\vect{b}} \cdot \mathbb{D}^{-1}_{\vect{k}} \cdot \vect{x} + \nonumber \\
      & & \kern 1.7cm + \left( \mathbb{D}^{-1}_{\vect{k}} \cdot \mathbb{D}_{\vect{x}} \cdot \vect{H} \right)^2 \cdot \mathbb{D}_{\vect{b}} \cdot \mathbb{D}^{-1}_{\vect{k}} \cdot \vect{x} + \nonumber \\
      & & \kern 1.7cm + \left( \mathbb{D}^{-1}_{\vect{k}} \cdot \mathbb{D}_{\vect{x}} \cdot \vect{H} \right)^3 \cdot \mathbb{D}_{\vect{b}} \cdot \mathbb{D}^{-1}_{\vect{k}} \cdot \vect{x} + \cdots \nonumber \\[0.4cm]
      \vect{e} &\approx& \left( \mathbb{D}_{\vect{b}} \cdot \mathbb{D}^{-1}_{\vect{k}} + \mathbb{D}^{-1}_{\vect{k}} \cdot \mathbb{D}_{\vect{x}} \cdot \vect{H} \cdot \mathbb{D}_{\vect{b}} \cdot \mathbb{D}^{-1}_{\vect{k}} \right) \cdot \vect{x}
      \label{approx_invDN}
\end{eqnarray}


\blue{Second, in the case of the Wilson-Cowan model (Eq.~\ref{invWC}) we approximate the saturation function $f(\vect{x})$ so that we can isolate the vector $\vect{x}$.
This can be done by expressing $f(\vect{x})$ through an Euler integration of $n$ terms:
$f(\vect{x}) = f(\vect{0}) + \int_{\vect{0}}^{\vect{x}} \frac{df}{dx}(\vect{x'}) d\vect{x'} \approx f(\vect{0}) + \sum_{\beta=0}^{n-1}  \frac{df}{dx}(\frac{\beta}{n}\vect{x}) \frac{\vect{x}}{n}$. Note that along the integration the derivatives are computed at different points from $\vect{0}$ up to $\frac{n-1}{n}\vect{x}$. If $n=1$ we have the (in principle poor) Maclaurin approximation and if $n \rightarrow \infty$ the result is perfect. In between, for finite $n$, we have an approximation with certain accuracy. In this case, taking into account that in the activation functions $f(\vect{0}) = \vect{0}$, and calling $g_n(\vect{x}) = \frac{1}{n}\sum_{\beta=0}^{n-1} \frac{df}{dx}(\frac{\beta}{n}\vect{x})$, we can write:}
\begin{equation}
      \blue{\vect{e} \approx \left( \mathbb{D}_{\vect{\alpha}} + \vect{W} \cdot \mathbb{D}_{g_n(\vect{x})} \right) \cdot \vect{x}}
      \label{invWC2}
\end{equation}


%

Now, the identification between the approximated versions of the decoding equations, Eq.~\ref{approx_invDN} and Eq.~\ref{invWC2},
is straightforward. As a result, we get the following relations between the parameters of both models:
\begin{eqnarray}
      \vect{\alpha} &=& \frac{\vect{b}}{\vect{k}} \nonumber \\
      \vect{W} &=& \mathbb{D}_{\left(\frac{\vect{x}}{\vect{k}}\right)} \cdot \vect{H} \cdot \mathbb{D}^{-1}_{\left(\frac{\vect{k}}{\vect{b}} \odot \blue{g_n(\vect{x})}\right)}
      \label{relation_W_H}
\end{eqnarray}

\noindent where the symbol $\odot$ denotes the element-wise (or Hadamard) product, and the ratios between vectors are also Hadamard divisions. 

Note that the Divisive Normalization kernel which is compatible with Eq.~\ref{relation_W_H}, $\vect{H} = \mathbb{D}_{\left(\frac{\vect{k}}{\vect{x}}\right)} \cdot \vect{W} \cdot \mathbb{D}_{\left(\frac{\vect{k}}{\vect{b}} \odot \blue{g_n(\vect{x})}\right)}$, has exactly the same structure as the one in Eq.~\ref{new_kernel_eq}.
Therefore, both models \blue{agree} if the Divisive Normalization kernel \emph{inherits} the structure from the Wilson-Cowan kernel left- and right-multiplied by these diagonal matrices, $\mathbb{D}_{\left(\frac{\vect{k}}{\vect{x}}\right)}$ and $\mathbb{D}_{\left(\frac{\vect{k}}{\vect{b}} \odot \blue{g_n(\vect{x})}\right)}$, respectively.

This theoretical result suggests an explanation for the structure that had to be introduced \emph{ad-hoc} in \cite{Martinez19} just to reproduce contrast masking.
Note that the interaction in the Wilson-Cowan case may be understood as wiring between sensors tuned to similar features, so a unit-norm Gaussian, $\vect{W} = \vect{H}^{\vect{ws}}$, is a reasonable choice \cite{Wilson73,Faugueras09}.
Note also that the weights before and after $\vect{W}$ (the diagonal matrices) are signal dependent.
Therefore, a fixed wiring $\vect{W}$ implies that the kernel in Divisive Normalization should be \emph{adaptive}. The one in the left, $\mathbb{D}_{\left(\frac{\vect{k}}{\vect{x}}\right)}$, has a direct dependence on the inverse of the signal, while the one in the right, $\mathbb{D}_{\left(\frac{\vect{k}}{\vect{b}} \odot \blue{g_n(\vect{x})}\right)}$, depends on the derivatives of the activation $f(\vect{x})$.
Next Section shows that these vectors $\frac{\vect{k}}{\vect{x}}$ and $\frac{\vect{k}}{\vect{b}} \odot \blue{g_n(\vect{x})}$ do have the high-pass frequency nature that explains why the low frequencies in $\vect{e}$ had to be attenuated \emph{ad-hoc} by introducing $\mathbb{D}_{\vect{l}}$ and $\mathbb{D}_{\vect{r}}$. We also show that the term of the right, $\mathbb{D}_{\left(\frac{\vect{k}}{\vect{b}} \odot \blue{g_n(\vect{x})}\right)}$, produces the shape changes needed on the interactions.

It is important to stress that the simplifications made in the decoding equations to get the analytical relations in Eq.~\ref{relation_W_H}
were done only for the sake of simplicity in the final relations obtained.
In summary, the expressions in Eq. \ref{relation_W_H} are exact for the simplified versions of the models. Considering the full version of the models, Eq. \ref{relation_W_H} would be an approximation. However, the experiments  below point out the validity of this approximation, because: (a)~we explicitly check that the errors are small in a range of scenarios, and (b)~we check that plugging these expressions into the full versions of the models also leads to consistent results.

\section{\blue{Numerical Experiments}}

The analysis of the proposed relation between the Divisive Normalization (DN) and the Wilson-Cowan (WC) models is a three stage process.
\emph{First}, one should take biologically plausible parameters (either in DN, in WC, or in both) and then, use the proposed expressions to build versions of the models supposed to behave similarly.
\emph{Second}, one should check if the models obtained in that way actually behave similarly. So that finally, \emph{third}, one can elaborate on the consequences of this correspondence.


In this experimental analysis, in Section~\ref{Parameters}, we build a psychophysically-inspired Wilson-Cowan model for V1 from a Divisive Normalization with psychophysically-tuned parameters~\cite{Malo15,Martinez17,Martinez19}. This model also preserves the basic properties of the interaction kernel and the saturation function of the Wilson-Cowan literature~\cite{Wilson73,Bressloff03,Faugueras09}. This Wilson-Cowan model should behave similarly to the corresponding Divisive Normalization model.

Then, Section \ref{math_properties} \blue{experimentally checks the mathematical relation} between the models.
In particular, \blue{for a wide range of parameters}:
(a) we show that the integration of the Wilson-Cowan equation certainly converges to a solution which is close to its corresponding Divisive Normalization; 
\blue{(b) we quantify the accuracy of the approximations required to get the relation between the models;} and
(c) we show that the Divisive Normalization solution is a stable node of the dynamical system governed by the Wilson-Cowan equations.

Finally, in Section \ref{consequences}, we address different \emph{consequences on contrast perception} using the proposed relation:
(a)~we analyze the signal-dependent behavior of the theoretically derived kernel and the benefits of the high-pass behavior to moderate the weight of the low-frequency components;
(b)~we show that the shape of the interactions between sensors changes depending on the surround;
(c)~we reproduce the contrast response curves with the proposed signal-dependent kernel; and
(d)~we discuss the use of the derived kernel in predicting the subjective metric of the image space.

\subsection{Psychophysically plausible parameters for a Wilson-Cowan model in V1}
\label{Parameters}


A possible way to check the  \blue{relation} between the models in V1 consists of starting
from the (lower-level / mechanistic / physiological) Wilson-Cowan model and let it
evolve to see if it converges to the (psychophysical) Divisive Normalization response.
To this end, for our Wilson-Cowan model, we need reasonable $\vect{\alpha}$, $\vect{W}$, and $f(\vect{x})$, for $\vect{e}$ and $\vect{x}$ defined in certain wavelet representation.

For the wavelet representation here we assume 4-orientation steerable transforms~\cite{Simoncelli92} as a convenient model of the simple cells (as done in~\cite{Schwartz01,Martinez17,Martinez19}).
In the experiments involving the (computationally intensive) integration of the Wilson-Cowan differential equation, Section~\ref{math_properties}, we used wavelets with 3~scales in $40\times40$ images to speed up the computation. But in the psychophysical illustrations, Section~\ref{consequences}, we used 4~scales in $64\times 64$ images. Of course these choices have to be considered as illustrative options. The selected model~\cite{Martinez17,Martinez19}, has the elements of previous image-computable models~\cite{Watson97,Winawer13,Schutt17}, but of course, the different instances differ in 
details despite they have the same qualitative flavour.  

The reference parameters for the nonlinearity are taken from the Divisive Normalization model in~\cite{Martinez19}. In that case, the parameters corresponding to contrast computation, contrast sensitivity, and masking in the spatial domain were directly measured using \emph{Maximum Differentiation} psychophysics~\cite{Malo15}, while the parameters related to brightness and to masking in the wavelet domain were tuned to reproduce subjective image quality data~\cite{Martinez17} and contrast perception curves~\cite{Martinez19}.

\begin{figure}[t!]
 \begin{centering}
 \hspace{-1cm}\includegraphics[height=11.5cm]{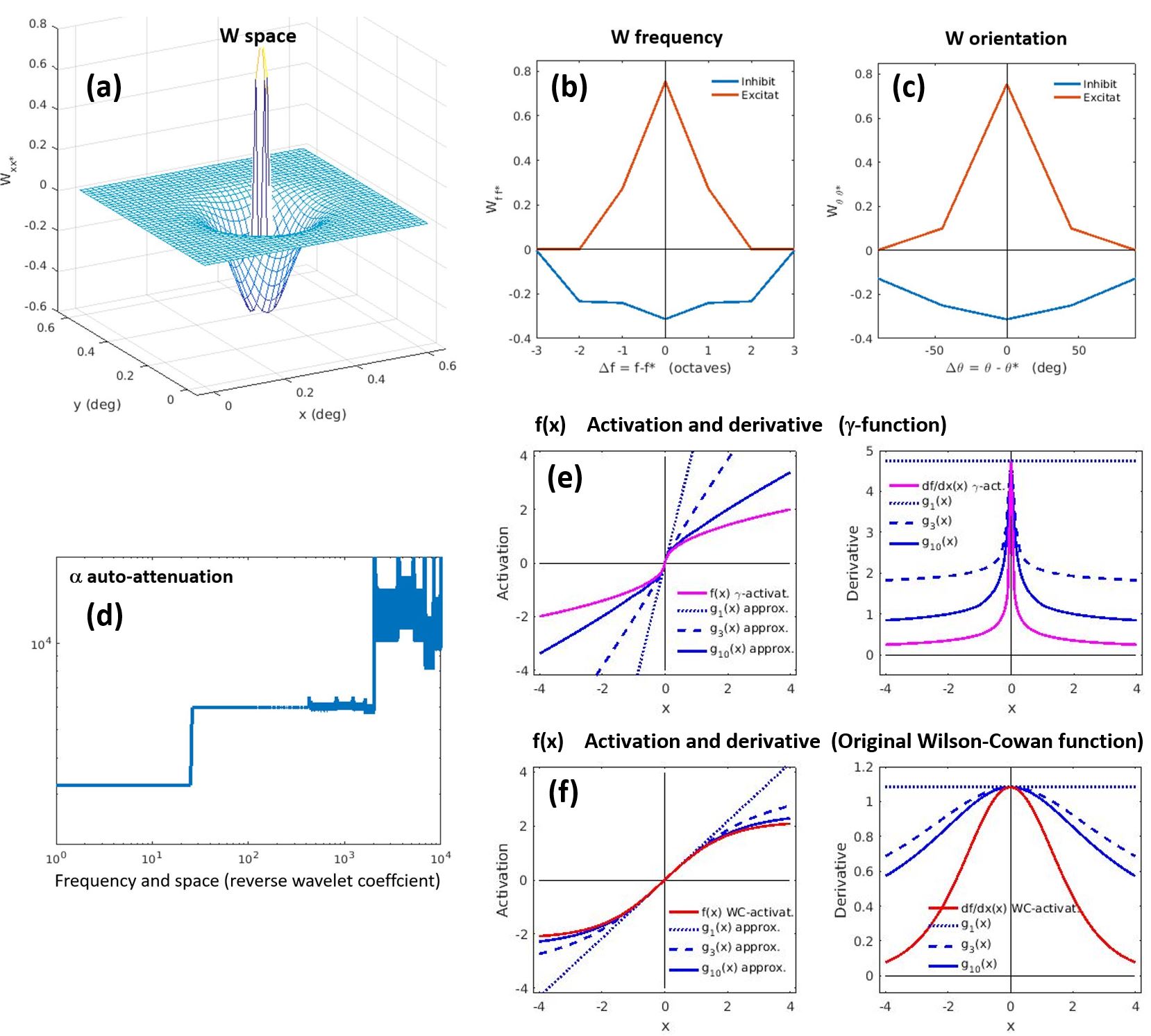}
 \end{centering}
 \captionof{figure}{\small{\textbf{Psychophysically-inspired parameters for Wilson-Cowan model.}
 Illustration of the Gaussian neighborhoods $\vect{W}$ with excitatory and inhibitory parts: separable interactions depending on departure in space \textbf{(a)}, frequency \textbf{(b)}, and orientation \textbf{(c)}.
The spatial component of $\vect{W}$ is shown for the central location of the considered visual field and the 24 cpd vertical subband.
Following Watson and Solomon~\cite{Watson97} the Gaussian kernels (here difference of Gaussians) are separable in space, frequency and orientation. Therefore lower frequency subbands have coarser sampling (and thus higher amplitude) but the same shape in space.
The shape is also the same for subbands tuned to different orientations. Equivalent separability applies for variations in frequency and orientation: the interactions in the plots (b) and (c) are the same for every spatial location (and orientation and frequency respectively).
The above plots display the (more intuitive) $-\mathbf{W}$, where positive and negative values mean excitatory and negative interaction respectively, but note that, according to the sign in Eq.~\ref{EqWC2}, positive weights are inhibitory.
\textbf{(d)} Auto-attenuation factor~$\vect{\alpha}$. In this plot the horizontal axis (wavelet coefficients in reverse order) can be qualitatively understood as \emph{frequency} so high frequencies display bigger attenuation. \blue{The panels \textbf{(e)} and \textbf{(f)} show different options for the pointwise activation nonlinearity, $f(\vect{x})$.
In pink we have the $\gamma$-function proposed in~\cite{Martinez17}
and in red we have the original proposal of Wilson-Cowan~\cite{Wilson73}. In both cases we show the approximations of these functions using $g_n(\vect{x})$ (for increasing number of terms $n$), and the corresponding derivatives, which have an impact in the relation between the kernels in the different models (Eq.~\ref{relation_W_H}).}}}
\label{param}
\end{figure}

As stated after Eq.~\ref{relation_W_H}, we took $\vect{W}$ as a Watson-Solomon separable Gaussian kernel~\cite{Watson97} with widths in space/frequency/orientation taken from the psychophysically plausible values in~\cite{Martinez19}.
In order to include both excitatory and inhibitory populations we complemented this initial kernel
with narrow excitatory neighborhoods whose width was a fraction of the original inhibitory neighborhoods. Finally we normalized the absolute amplitude of the neighborhoods to have unit-norm center-surround interactions.
Figures~\ref{param}.a-c illustrate the psychophysically sensible separable kernels $\vect{W}$. These unit-norm kernels $\vect{W}$ scaled as in~\cite{Watson97,Martinez19} are consistent with the shapes used in the Wilson-Cowan literature~\cite{Bressloff03,Faugueras09}.

Regarding the auto-attenuation we simply took the constants $\vect{k}$ and $\vect{b}$ from~\cite{Martinez19} and
used the first equation of the proposed relation, Eq.~\ref{relation_W_H}, to obtain~$\vect{\alpha}$.
Fig.~\ref{param}.d shows the $\vect{\alpha}$ vector for the 3-scale wavelet (coefficients ordered from low-to-high frequency). Note that the response of sensors tuned to higher frequencies is more attenuated in the evolution of the differential equation while low frequencies have lower auto-attenuation.

\blue{Finally, Figs.~\ref{param}.e and~\ref{param}.f, display different activations $f(\vect{x})$ that we used in the experiments together with representative Euler approximations, $g_n(\vect{x}) \, \vect{x}$, and the functions related to their derivatives, $\frac{df}{dx}(\vect{x})$ and $g_n(\vect{x})$. These activation functions include the \emph{original}-activation in~\cite{Wilson73,Bressloff03}, and the so called $\gamma$-activation inspired in retinal transduction~\cite{Martinez17}.
Appendix~\ref{activations} gives the expressions of these activation functions.}
In our wavelet case, the horizontal and vertical axes of the function $f(\cdot)$ to be applied to each coefficient $x$ of certain subband are scaled by the average amplitude of the responses of the corresponding linear sensors to natural images. 
With that scaling the nonlinearities preserve the relative scales of the input subbands in the vector $\vect{e}$ that comes from the linear filters.

\blue{In the next experiments the above psychophysically sensible parameters (the reference values) are modified in several ways to show that the proposed relation works for a wide range of model parameterizations. Specifically, 
(a)~we explored different widths of the interaction kernels 
by using five scaling factors applied to the reference widths: from unrealistically narrow (zero-width, identity kernels that disregard interactions), to unrealistically wide kernels (where the reference widths are increased by an order of magnitude); (b)~we considered kernels with the above mentioned excitatory-inhibitory nature, and kernels with just-inhibitory nature; and (c)~we considered two possible activations (the original-activation and the $\gamma$-activation). We considered a total of 12 parameterizations of the models: 5~kernel widths $\times$ 2~excit-inhib configurations $\times$ 1~activation (original-activ.) $+$ 1~kernel width (the reference one) $\times$ 2~excit-inhib configurations $\times$ 1~activation (the $\gamma$-activ.).}

The interested reader has access to the specific values of the parameters in Appendix~\ref{activations} for the activations, and in the code that reproduces all the simulations of the paper (described in  Appendix~\ref{code}). 


\subsection{\blue{Experimental check of mathematical properties}}
\label{math_properties}

\subsubsection{Wilson-Cowan systems converge to the Divisive Normalization}

The Wilson-Cowan expression, Eq. \ref{EqWC2}, defines an initial value problem where the response at time zero evolves (or is updated) according to the right hand side of the differential equation.
In our case, we assume that the initial value of the output is just the input $\vect{x}(0) = \vect{e}$. Moreover, as we deal with \blue{static} images, we assume that the input is constant. And then, we solve this first \blue{order} differential equation by the simplest (Euler) integration method:
\vspace{-0.1cm}
\begin{eqnarray}
      \vect{x}(t+\blue{\Delta t}) = \vect{x}(t) + \Bigl( \vect{e} - \mathbb{D}_{\vect{\alpha}} \cdot \vect{x}(t) - \vect{W} \cdot f(\vect{x}(t)) \Bigr) \blue{\Delta t}
      \label{initial_solution_Euler}
\end{eqnarray}

\begin{figure}[!t]
 \centering
 \begin{tabular}{c}
 \hspace{-1cm}\includegraphics[height=0.53\linewidth]{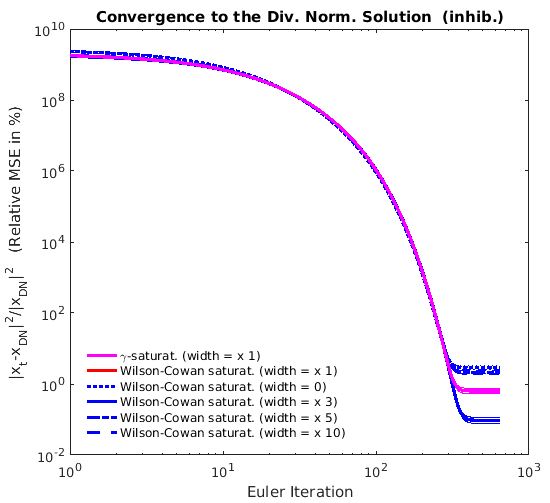}\hspace{0.25cm}\includegraphics[height=0.53\linewidth]{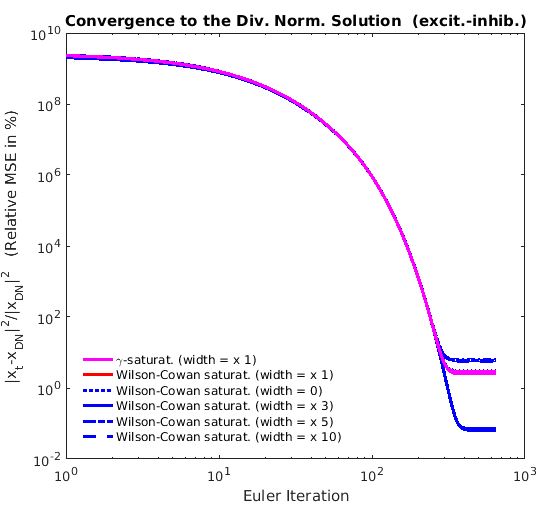}\\[-0.0cm]
 \end{tabular}
 \captionof{figure}{\small{\textbf{Convergence to the Divisive Normalization solution (I): quantitative error.} 
 \blue{Percentage of relative MSE between the solution of the Wilson-Cowan equations and the Divisive Normalization response as a function of discrete time (steps in Euler integration).  
 \emph{Left}: relative MSE for different kernel widths and activation/saturation functions in case of just inhibitory interactions. \emph{Right}: relative MSE for different kernel widths and saturation functions in case of excitatory+inhibitory interactions.
 The configurations in pink and red are the psychophysically meaningful ones, which achieve 0.6\% and 2.7\% relative MSE respectively. In all cases, the update of the solution tends to zero (results not shown) indicating the approach to a steady state. The curves in bold style represent the median (50 quantile) over 45 natural images, and the curves in light style (very close to the median), represent the 25 and 75 quantiles. This result suggest the succesful convergence of WC to DN for natural images over a wide range of model parameterizations.}}}
 \label{fig_conv}
\end{figure}

Figure \ref{fig_conv} shows the evolution of the response obtained from this integration, applied to 45 natural images taken from calibrated databases~\cite{VanHateren98,Laparra12}, \blue{using the biologically sensible parameters $\vect{\alpha}$, $\vect{W}$ and $f(\cdot)$ presented in Fig.~\ref{param}~\cite{Malo15,Martinez17,Martinez19,Wilson73,Bressloff03,Faugueras09}, 
and the mentioned variations to cover a wide range of model parameterizations}. 
Our Euler integration used a small enough discrete time step, $\blue{\Delta t}=10^{-5}$, and the initial responses, the vectors $\vect{e}$, were computed using the first 3~layers of the model in Fig.~1~\cite{Martinez17,Martinez19} followed by a linear steerable wavelet of 4 orientations and 3 scales.
\blue{The integration requires no approximation of the WC model, and the evolving solution is checked against the corresponding DN response that uses the proposed Eq.~\ref{relation_W_H}.}

As can be seen, \blue{the solution of the Wilson-Cowan integration converges to the Divisive Normalization solution because their difference (percentage of relative Mean Squared Error) decreases as it evolves in \emph{all} the 12 considered parameterizations.
The relative MSE in the psychophysically meaningful situations ($\times$1 width) is below $3\%$ (lines in pink and red), and for the other configurations it is always below $6\%$. Therefore, the Divisive Normalization always explain more than $94\%$ of the energy of the Wilson-Cowan solution.}
Moreover, these results represent the steady states because the updates of the solutions in the integrals always tend to zero (results not shown).


\begin{figure}[!t]
 \centering
 \begin{tabular}{c}
 \hspace{-1cm}\includegraphics[height=0.85\linewidth]{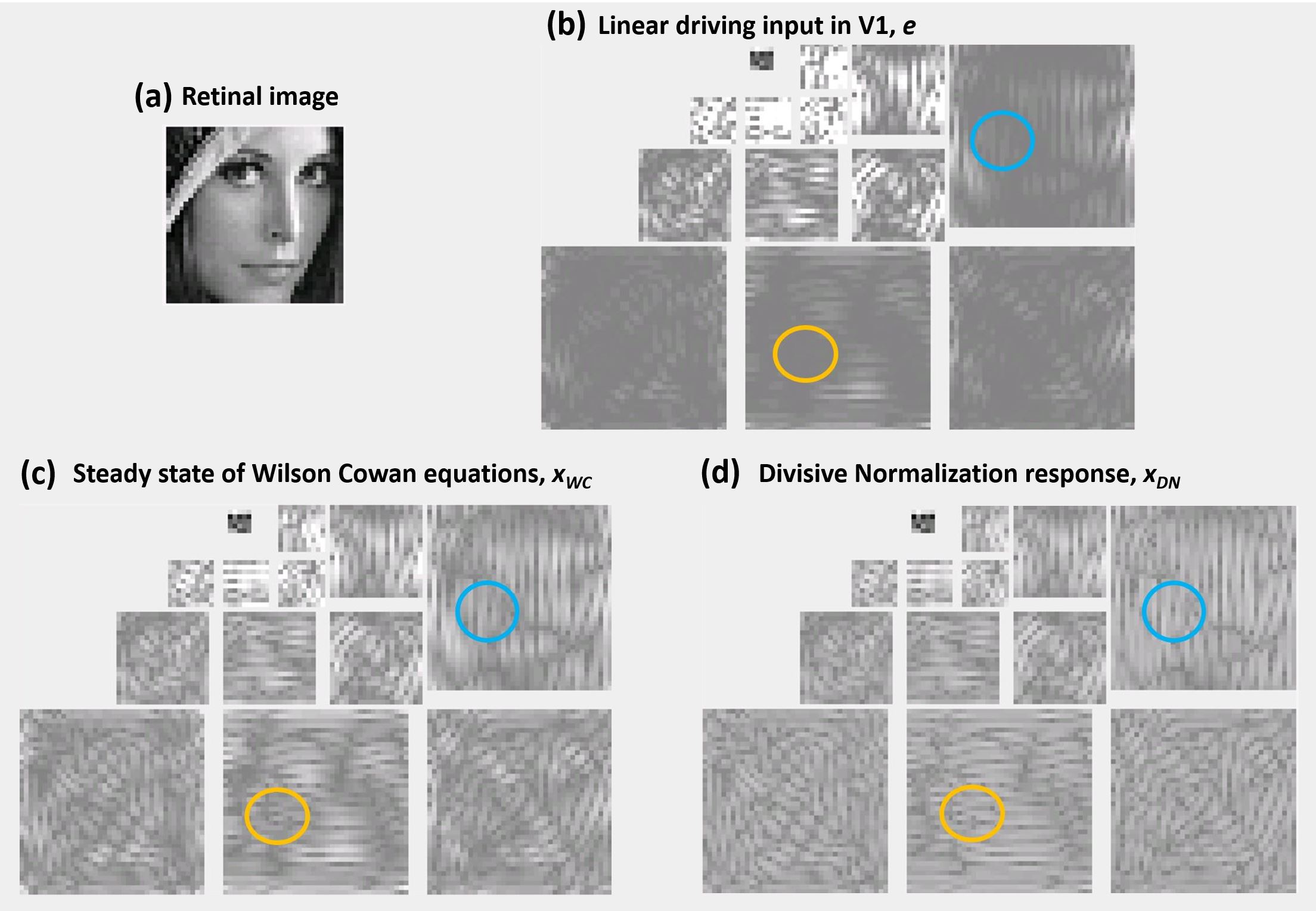}\\[-0.0cm]
 \end{tabular}
 \captionof{figure}{\small{\textbf{Convergence to the Divisive Normalization solution (II): qualitative similarity.}
 \textbf{(a)} Input image (stimulus at the retina), \textbf{(b)} Responses of the linear simple cells of V1. Responses are spatially non-stationary, see regions of very low amplitude highlighted in yellow and blue. \textbf{(c)} Steady state of the Wilson-Cowan equation after 650 iterations of Eq.~\ref{initial_solution_Euler}
 with inhibitory kernel $W$ of psychophysically sensible width and $\gamma$-activation, \textbf{(d)}  Corresponding Divisive Normalization response using the kernel given by Eq.~\ref{new_kernel_eq}. \blue{In this example the relative MSE between $\vect{x}_{\text{WC}}$ and $\vect{x}_{\text{DN}}$ is 0.93\%, but, more importantly, note how the amplitude of the response of the highlighted neurons has increased similarly in the nonlinear cases leading to a more stationary response.}}}
 \label{illustr_converg}
\end{figure}

Fig.~\ref{illustr_converg} illustrates the qualitative similarity of the responses of the two models
and their \blue{comparable} equalization effect in the wavelet domain.
The nonlinear response $\vect{x}_{\textrm{WC}}$ was computed by integrating Eq.~\ref{initial_solution_Euler}, and $\vect{x}_{\textrm{DN}}$ was computed with Eq.~\ref{DN_B}.
We used the parameters introduced in Section~\ref{Parameters} and the corresponding parameters for Divisive Normalization using Eq.~\ref{relation_W_H}.

Note how the nonlinearities substantially increase the amplitude of the signal in the regions where the linear response is low.
The regions highlighted in blue and orange in $\vect{e}$ display low activity compared to their neighbors because there are no edges in those regions of the image.
However, the corresponding neurons after Wilson-Cowan or Normalization have increased their activity.
The amplitude of the signal after the nonlinearities is more stationary across the subbands.
Moreover, the nonlinearities lead to responses where the image structure is less apparent:
the activity of a neuron is more independent from the activity of the neighbors.
Equalization and increased independence qualitatively suggested in Fig.~\ref{illustr_converg}
are consistent with previous (quantitative) studies that report redundancy reduction both in Divisive Normalization~\cite{Schwartz01,Malo10,Malo20}, and in the Wilson-Cowan model~\cite{Gomez19}.

\blue{\subsubsection{Quantification of the accuracy of the approximations}}

\noindent 
\blue{The proposed relation, Eq.~\ref{relation_W_H}, is based on two approximations:}
\begin{itemize}
\item \blue{The approximation of the \emph{inverse} of Divisive Normalization 
to obtain Eq.~\ref{approx_invDN}, namely:
$\left( I - \mathbb{D}^{-1}_{\vect{k}} \cdot \mathbb{D}_{\vect{x}} \cdot \vect{H} \right)^{-1} \approx I + \mathbb{D}^{-1}_{\vect{k}} \cdot \mathbb{D}_{\vect{x}} \cdot \vect{H} $.}  
\item \blue{The approximation of $f(\vect{x})$ in Wilson-Cowan to obtain Eq.~\ref{invWC2}, namely:
$f(\vect{x}) \approx g_n(\vect{x}) \, \vect{x} = \left( \frac{1}{n}\sum_{\beta=0}^{n-1} \frac{df}{dx}(\beta \frac{\vect{x}}{n}) \right) \vect{x}$.}
\end{itemize}

\blue{The accuracy of such approximations depends on the model parameters, e.g. the shape and magnitude of $\vect{H}$ or $f(\cdot)$, and on the responses $\vect{x}$ to natural images. 
The low amplitude of the coefficients of natural images in wavelet representations~\cite{Olshausen96,malo2000role}, and the accuracy of similar approximations for psychophysically sensible parameters~\cite{Malo06a} suggest that errors will be small. However, in this section we explicitly compute both sides of the above expressions (with and without approximation) for a range of representative images and model parameterizations, and we compute the difference between both sides.
This difference is the error due to the approximation.
We express the energy of the difference (the Mean Squared Error) in terms of percentage of the energy of the function with no approximation: the relative MSE in \%.} 

\blue{In Table 1~\label{relative_error_table} we show the relative MSE (in \%) for both approximations (\emph{inverse} and \emph{f}(\textbf{x})) together with the error in \emph{convergence}, also in relative MSE, for 12 different model parameterizations. The approximations of $f(\vect{x})$ were done using $g_{10}(\vect{x})$, i.e. computing derivatives in 10 points.} 

\blue{The approximations generally explain more than $90\%$ of the energy of the original magnitudes. The only exception is the approximation of the inverse of the divisive normalization for the (unrealistic) zero-width kernel, where the relative MSE amounts to $\sim30\%$. The deviation in this unrealistic case makes sense because reducing the width of unit-volume kernels increases their height and hence the magnitude of $\vect{H}$ increases so the term summed to the identity in the expression under consideration is not as small. This leads to an increased error in the approximation.} 

\blue{Interestingly, in the whole range of parameterizations considered, the approximations do not have a big impact in the convergence error, which is the actual measure of correspondence between the two models.}

\begin{table}[b]
\centering
\hspace{-1.5cm}
\resizebox{15cm}{!}{%
\begin{tabular}{|l|ccc|ccc|}
\cline{2-7}
\multicolumn{1}{c|}{} & \multicolumn{3}{c|}{\cellcolor{gray!30}\textbf{Inhibitory W}} & \multicolumn{3}{c|}{\cellcolor{gray!30}\textbf{Excitatory-Inhibitory W}} \\ \cline{2-7}
\multicolumn{1}{c|}{} & \cellcolor{gray!30}\textbf{Inverse} & \cellcolor{gray!30}$\emph{f}(\textbf{x})$ & \cellcolor{gray!30}\textbf{Convergence} & \cellcolor{gray!30}\textbf{Inverse} & \cellcolor{gray!30}$\emph{f}(\textbf{x})$ & \cellcolor{gray!30}\textbf{Convergence} \\ \hline
\multicolumn{7}{|c|}{\cellcolor{gray!90}\hspace{1.6cm}$\mathbf{\gamma}$-\textbf{activation}} \\ \hline
\textbf{Width} $\times$1 & 0.019 $\pm$ 0.009 & 0.0001 $\pm$ 0.0002 & \textcolor{red}{\textbf{0.7 $\pm$ 0.1}} & 10 $\pm$ 6 & {0.0005 $\pm$ 0.0001} & \textcolor{red}{\textbf{2.6 $\pm$ 0.3}} \\ \hline
\multicolumn{7}{|c|}{\cellcolor{gray!90}\hspace{0.65cm}$\emph{Original}$-\textbf{activation}} \\ \hline
$\times$0 & 32.74 $\pm$ 0.04 & {0.07 $\pm$ 0.01} & 2.9 $\pm$ 0.3 & 31.04 $\pm$ 0.05 & {0.06 $\pm$ 0.02} & 2.9 $\pm$ 0.3 \\ \hline
$\times$1 & 0.019 $\pm$ 0.009 & {0.04 $\pm$ 0.01} & \textcolor{red}{\textbf{0.6 $\pm$ 0.1}} & 10 $\pm$ 6 & {0.04 $\pm$ 0.03} & \textcolor{red}{\textbf{2.7 $\pm$ 0.4}} \\ \hline
$\times$3 & 0.00022 $\pm$ 0.00004 & {0.11 $\pm$ 0.02} & 0.09 $\pm$ 0.01 & 0.00014 $\pm$ 0.00002 & {0.06 $\pm$ 0.04} & 0.067 $\pm$ 0.007 \\ \hline
$\times$5 & $6.0 \times 10^{-8} \pm 0.3 \times 10^{-8}$ & {0.13 $\pm$ 0.02} & 2.1 $\pm$ 0.2 & $1.1 \times 10^{-7} \pm 0.6 \times 10^{-7}$ & {0.05 $\pm$ 0.01} & 6.0 $\pm$ 0.5 \\ \hline
$\times$10 & $1.6 \times 10^{-10} \pm 0.1 \times 10^{-10}$ & {0.009 $\pm$ 0.005} & 2.1 $\pm$ 0.2 & $1.5 \times 10^{-8} \pm 0.1 \times 10^{-8}$ & {0.10 $\pm$ 0.03} & 5.8 $\pm$ 0.5 \\ \hline
\end{tabular}}
\vspace{0.2cm}
\caption{\blue{\small{\textbf{Accuracy of the approximations and convergence error.}
Relative Mean Squared Error (rel. MSE in \%) of three magnitudes: 
(a)~the error in the approximation of the \emph{inverse} of Divisive Normalization,
(b)~the error in the approximation of $f(\textbf{x})$ in Wilson-Cowan, and
(c)~the final error in the \emph{convergence} shown in Fig.~\ref{fig_conv}.
The convergence errors for the psychophysically sensible configurations are highlighted in bold style and color. Values represent the median relative MSE in \% over 45 images, and uncertainty intervals represent the 50\% interquantile distance.
}}}
\label{relative_error_table}
\end{table}

\subsubsection{Stability analysis of the Divisive Normalization response}

The stability of a dynamical system at the steady state is determined by the Jacobian with regard to perturbations in the response: if the eigenvalues of this Jacobian are all negative for this response, it is a stable node of the system~\cite{Logan15}. In that situation the evolution of the perturbations is a vector field oriented towards the stable node.

\begin{figure}[b]
 \centering
 \begin{tabular}{c}
 \hspace{-1cm}\includegraphics[width=0.55\linewidth,height=0.55\linewidth]{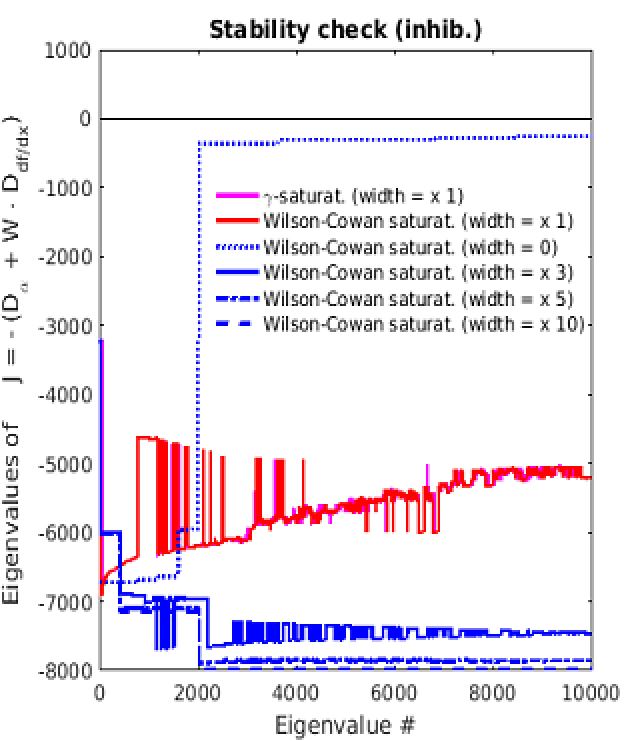}\hspace{0.2cm}
 \hspace{-0cm}\includegraphics[width=0.55\linewidth,height=0.55\linewidth]{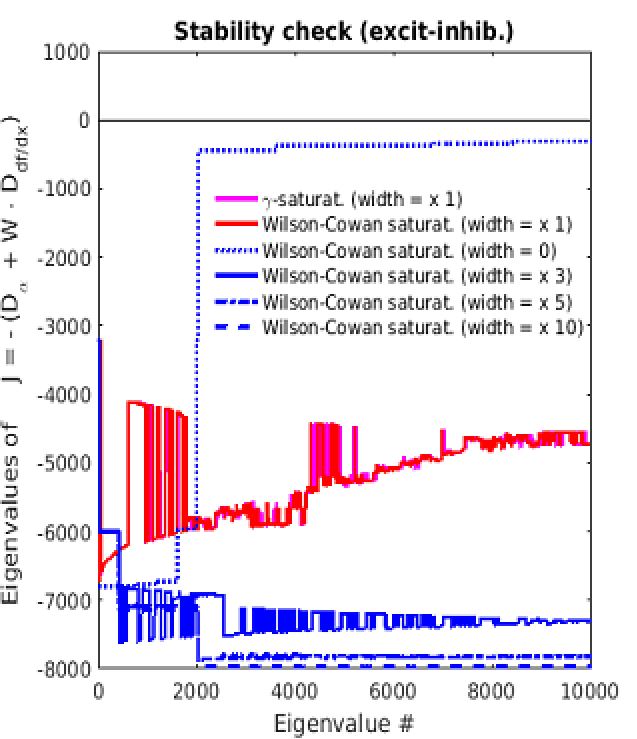}\\[-0.0cm]
 \end{tabular}
 \captionof{figure}{\small{\textbf{Stability of the Divisive Normalization solution (I).} Eigenspectrum of the Jacobian of the right-hand side of the Wilson-Cowan differential equation with psychophysically-tuned parameters on natural images \blue{(curves in pink and red), and different additional configurations (different widths and activation functions), including just inhibitory interactions (Left) and excitatory-inhibitory interaction (right)}. The curves refer to the median of the eigenvalues over $45$ representative images extracted from the calibrated datasets~\cite{VanHateren98,Laparra12}. The standard deviation and quartile distance is so small that cannot be seen in the plot. The result shows that the eigenvalues are \emph{all negative}. This suggests that the Divisive Normalization is a stable node of the system for natural images \blue{for a wide range of model parameterizations.}}}
 \label{fig_stab_eigen}
\end{figure}

In our case, the Jacobian with regard to the output signal of the right hand side of the Wilson-Cowan differential equation, Eq. \ref{EqWC2}, is:
\begin{equation}
   J = - (\mathbb{D}_{\alpha} + \vect{W} \cdot \mathbb{D}_{\frac{df}{dx}(\vect{x})})
   \label{eq_stabil}
\end{equation}
Fig.~\ref{fig_stab_eigen} shows the eigenvalues of this Jacobian using 
\blue{a wide range of parameters (the 12 configurations obtained through variations of the reference values presented in Section~\ref{Parameters})}, with responses from a set of 45 representative natural images from colorimetrically calibrated datasets~\cite{VanHateren98,Laparra12}.
This result shows that \emph{all} the eigenvalues are negative, thus suggesting that the Divisive Normalization solution is a stable node of the dynamical system. 

The stability of the system can be further illustrated by the visualization of the vector field of perturbations in the phase space of the system~\cite{Logan15}.
In this case we visualize this vector field for the Divisive Normalization solution.
As the signals in our problem live in very high-dimensional spaces (the wavelet vectors in this section have dimension 10025) it is not possible to visualize
the complete phase space, so we just select some illustrative 3-dimensional and 2-dimensional examples.

\begin{figure}[!t]
 \centering
 \begin{tabular}{cc}
 \hspace{-2cm}\includegraphics[height=0.65\linewidth]{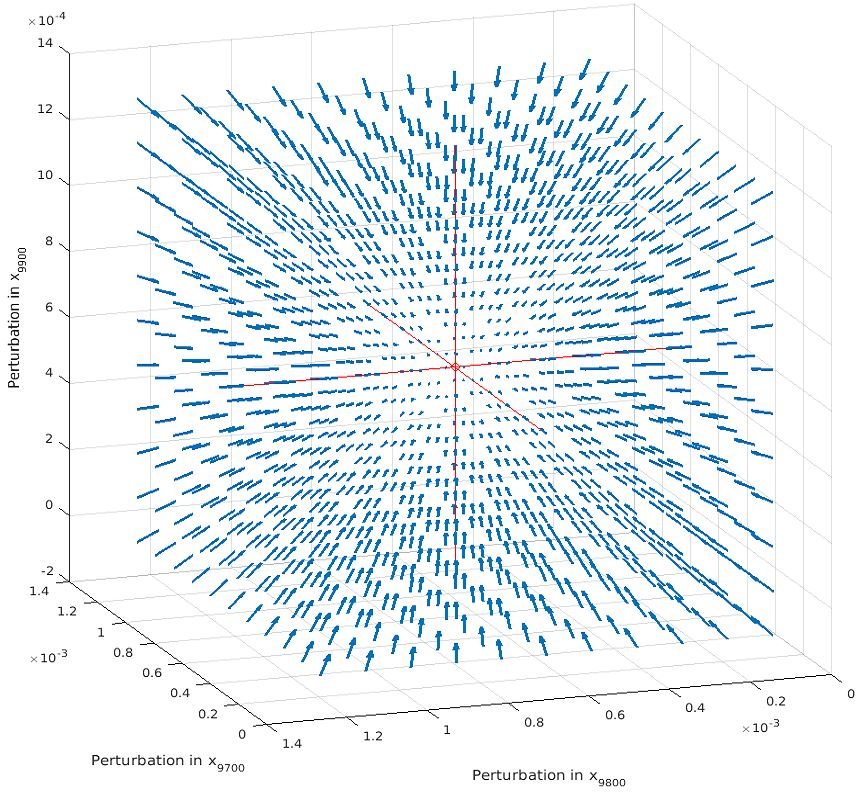} & \includegraphics[height=0.7\linewidth]{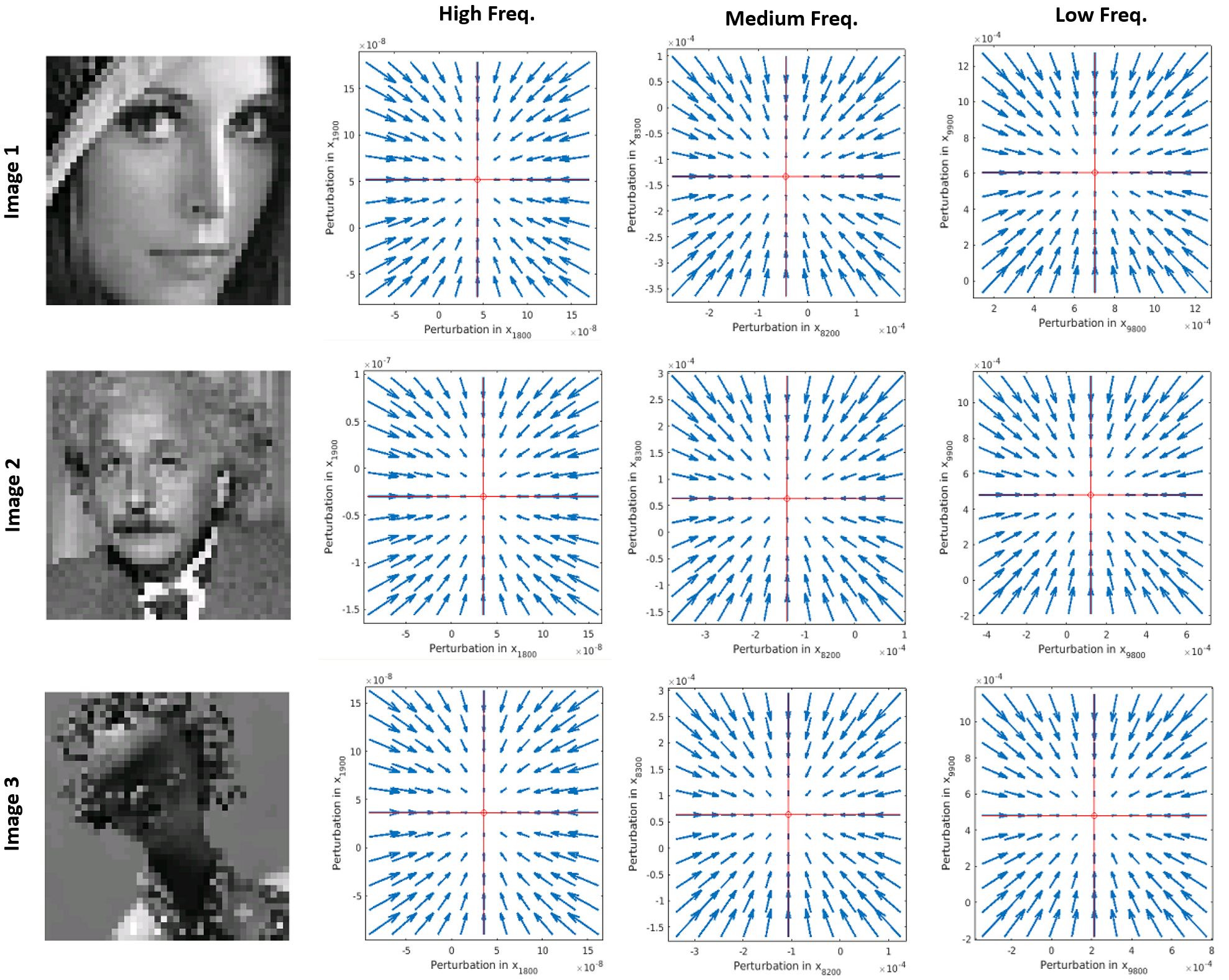} \\[-0.0cm]
 \end{tabular}
 \captionof{figure}{\small{\textbf{Stability of the Divisive Normalization solution (II).} Vector fields in the phase space generated by the Jacobian of the psychophysically-tuned Wilson-Cowan model. The example at the left (red dot) corresponds to the response of 3~low-frequency sensors at the Divisive Normalization solution. This vector field describes the evolution of the response if it is perturbed in arbitrary directions (the cardinal directions, in red, or any combination of them). The result is general: examples at the right show similar results for pairs of sensors tuned to different frequencies for different input stimuli.}}
 \label{fig_field_1}
\end{figure}

Fig.~\ref{fig_field_1} (left) shows an example taking just 3~neurons of the V1 layer.
In this case we took a particular image (the standard image \emph{Lena}) and
we focused on the response of 3~specific sensors of the low-frequency scale of the Divisive Normalization vector: the 9700, 9800 and 9900-th responses. In that way we get the red circle in Fig.~\ref{fig_field_1} (left).
Arbitrary perturbations of the responses of these neurons leads to the dynamics shown in the phase space: the vector field induced by the Jacobian implies that any perturbation is sent back to the original (no-perturbation) response, which is, then, a stable node of the system.

Similar behavior is obtained for coefficients of other subbands or other images. See Fig.~\ref{fig_field_1} (right), where, for simplicity, we consider perturbations in pairs of neurons.

\vspace{0.2cm}
In summary, the Divisive Normalization solutions are stable nodes of the \blue{corresponding} Wilson-Cowan systems. This conclusion confirms the assumption under the proposed relation: Divisive Normalization as a steady state of the Wilson-Cowan dynamics.

\subsection{Consequences on contrast perception}
\label{consequences}

The proposed relation implies that the Divisive Normalization kernel \emph{inherits} the structure of the Wilson-Cowan interaction matrix (typically Gaussian  \cite{Wilson73,Faugueras09}), modified by some specific signal dependent diagonal matrices, as seen after Eq. \ref{relation_W_H}, and allows to explain a range of contrast perception phenomena.

First, regarding the structure of the kernel,
we show that our prediction is consistent with previously required modifications of the Gaussian kernel in Divisive Normalization to reproduce contrast perception~\cite{Martinez19}.
Second, we show that the kernel in Divisive Normalization modifies its shape depending on the signal, thus explaining the behavior previously reported in \cite{Cavanaugh02b,Coen12}.
Third, we use the predicted signal-dependent kernel to simulate contrast response curves consistent with \cite{Foley94,Watson97}. And finally, the proposed relation is also applied to reproduce the experimental visibility of spatial patterns in more general contexts as subjective image quality assessment \cite{LIVE6,ponomarenko08,TID2008}.

In this section we do not integrate the Wilson-Cowan differential equation, but we use the expression for the steady state solution
with the kernel obtained from the proposed relation.
This alleviates computation so, as opposed to the previous Section, in the following examples we use a wavelet representation
of higher dimensionality, with 4 scales and 4 orientations, applied on bigger images, $64\times64$.
Regarding the parameters, we use unit-norm Gaussian kernels in $\vect{H}^{ws}$ or $\vect{W}$, and constants $\vect{k}$ and $\vect{b}$
also defined over 4~scales and 4~orientations, directly taken from~\cite{Martinez19}.

\subsubsection{Structure of the kernel in Divisive Normalization}
Here we compare the empirical filters $\mathbb{D}_{\vect{l}}$ and $\mathbb{D}_{\vect{r}}$, that had to be introduced \emph{ad-hoc} in \cite{Martinez19}, 
with the theoretical ones obtained through Eq. \ref{relation_W_H}.

\begin{figure}[!b]
	\centering
    \small
    \setlength{\tabcolsep}{2pt}
    \begin{tabular}{c}
    \hspace{-1cm} \includegraphics[width=1.3\textwidth]{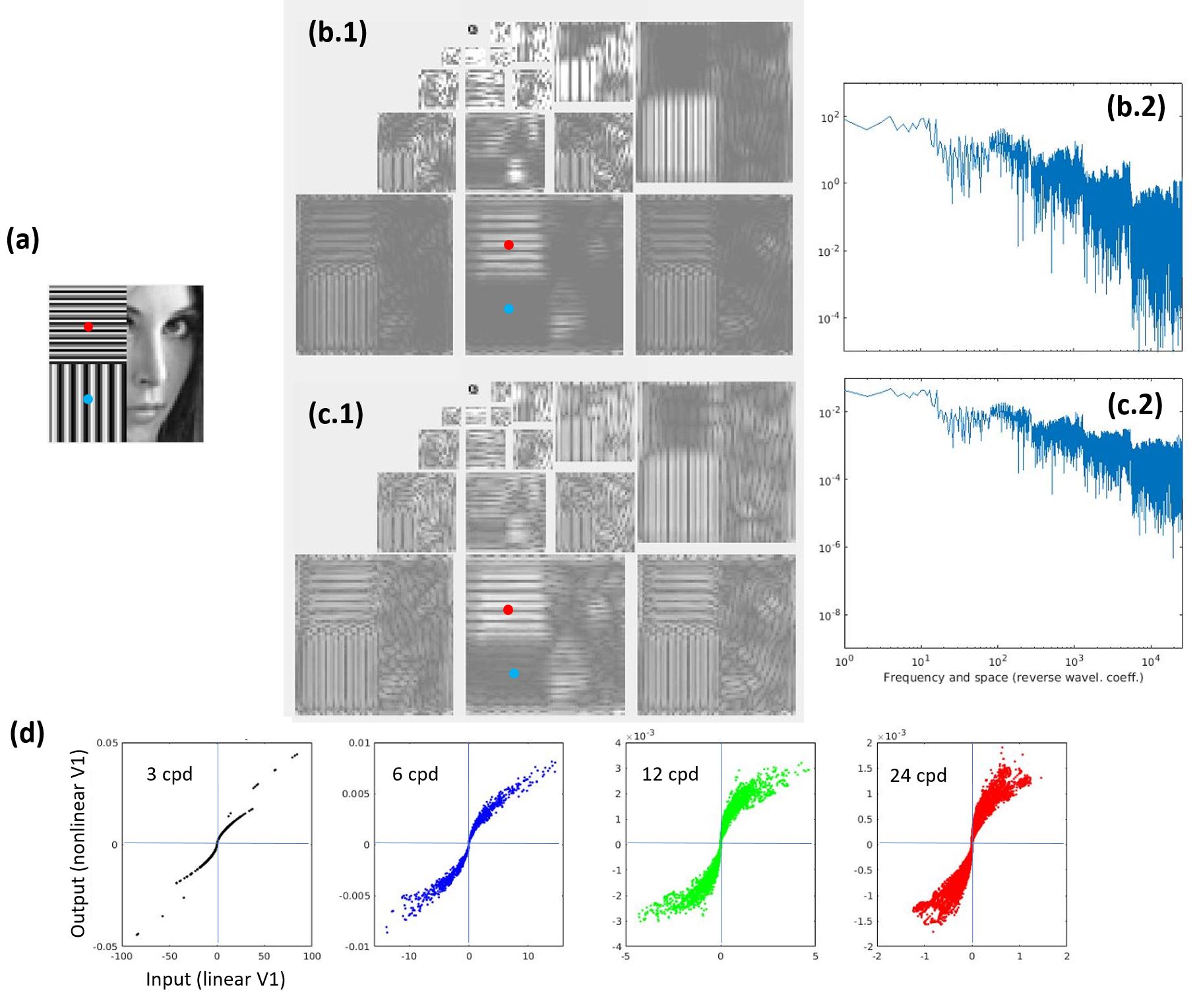} \\
    \end{tabular}
    \vspace{0.1cm}
	\caption{\small{\textbf{Linear and nonlinear 
    responses in V1 for an illustrative stimulus}.
     \textbf{(a)} Retinal image composed of a natural image and two synthetic patches of frequencies 24 and 12 cpd.
     This image goes through the first stages of the model (see Fig. 1) up to the cortical layer, where a set of linear wavelet filters leads to the responses with energy $\vect{e}$, which are nonlinearly transformed into the responses $\vect{x}$. \textbf{(b.1)} Wavelet panel that represents $\vect{e}$. 
     \textbf{(c.1)} Wavelet panel that represents $\vect{x}$. The highlighted sensors in red and blue (tuned to different locations of the 24 cpd scale, horizontal orientation) have characteristic responses given the image patterns in those locations. 
     The plots \textbf{(b.2)} and \textbf{(c.2)} show the vector representation of the wavelet responses arranged according to the MatlabPyrTools convention \cite{Simoncelli92}. These plots show how natural images typically have bigger energy in the low-frequency sensors. \textbf{(d)} Input-output scatter plots at different spatial frequencies, in cycles per degree (cpd), and demonstrate that Divisive Normalization (and the Wilson-Cowan solution) imply adaptive saturating nonlinearities depending on the neighbors (i.e. a family of sigmoid functions).}}\label{explanation}
    \vspace{0.3cm}
\end{figure}

Before going into the details of the kernel,
lets get some intuition on the typical structure of the vectors $\vect{x}$~and~$g_n(\vect{x})$.
Fig.~\ref{explanation} shows an illustrative stimulus with oriented textures and the corresponding responses of linear and nonlinear V1-like sensors based on steerable wavelets.
Typical responses for natural images are low-pass signals (see the vectors at Figs.~\ref{explanation}.b.2 and~\ref{explanation}.c.2).
The response in each subband is an adaptive (context dependent) nonlinear transduction
(Fig.~\ref{explanation}.d). Each point in Fig.~\ref{explanation}.d represents the input-output relation for each neuron in the subbands of the different scales (from coarse to fine). As each neuron has a different neighborhood, there is no simple input-output transduction function, but a scatter plot representing different instances of an adaptive transduction.

The considered image is designed to lead to specific excitations in certain sensors (subbands and locations in the wavelet domain).
Note, for instance, the high and low frequency synthetic patterns (24 and 12 cycles per degree, cpd, horizontal and vertical, respectively) in the image regions highlighted with the red and blue dots. In the wavelet representations we also highlighted some specific sensors in red and blue corresponding to the same spatial locations and the horizontal subband tuned to 24 cpd.
Given the tuning properties of the neurons highlighted in red and blue, it makes sense that wavelet sensor in red has bigger response than the sensor in blue.

\begin{figure}[!b]
	\centering
    \small
    \setlength{\tabcolsep}{2pt}
    \begin{tabular}{c}
    \hspace{-0.0cm} \includegraphics[width=0.95\textwidth]{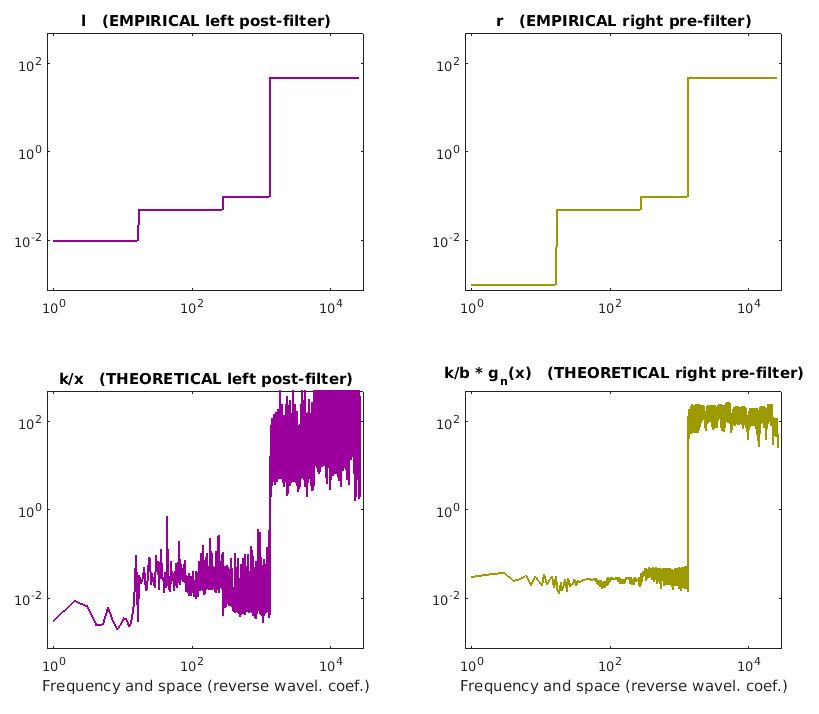} \\
    \end{tabular}
    \vspace{-0.15cm}
	\caption{\small{\textbf{Empirical and theoretical modulation of the Divisive Normalization kernel}. Vectors in diagonal matrices $D_{\vect{l}}$ and $D_{\vect{r}}$ that multiply the Gaussian kernel in the empirical tuning represented by Eq. \ref{new_kernel_eq} (top), and in the theoretically derived Eq.~\ref{relation_W_H} (bottom).
    \blue{These theoretical filters correspond to a specific natural image and using the $g_{10}(\vect{x})$ approximation of the $\gamma$-activation. The median relative RMSEs for the predicted filters over 45 natural images are 10.2\% (for the left filter) and 5.4\% (for the right filter). The difference for the right filter using the original Wilson-Cowan activation and also $g_{10}(\vect{x})$ is 5.1\% (curve not shown). However, as the empirical filters were just ad-hoc adjusted in~\cite{Martinez19}, here the relevant is the reproduction of the required high-pass structure, not the MSE.}}}
\label{Fig_filters}
    \vspace{0.1cm}
\end{figure}

With this knowledge of the signal in mind:
(1) low-pass trend in $\vect{x}$ shown in Fig. \ref{explanation},
(2) bigger derivative $g_n(\vect{x})$ for high frequencies because the derivative is higher for low amplitude signals (see Fig.~\ref{param}), and
(3) the vector $\vect{b}$ is bigger for low-frequencies \cite{Martinez19},
we can understand the high-pass nature of the vectors included in the diagonal matrices that appear at the left and right sides of the
theoretically-derived kernel $\vect{H} = \mathbb{D}_{\left(\frac{\vect{k}}{\vect{x}}\right)} \cdot \vect{W} \cdot \mathbb{D}_{\left(\frac{\vect{k}}{\vect{b}} \odot \blue{g_n(\vect{x})}\right)}$.

Fig. \ref{Fig_filters} compares the empirical \emph{left} and \emph{right} vectors, $\vect{l}$ and $\vect{r}$ that were adjusted \emph{ad-hoc} to reproduce contrast curves in \cite{Martinez19}, with those based on the theoretical relation proposed here. 
\blue{In this case we only consider the comparison with the psychophysically sensible parameterization since the ad-hoc tuning was done for that specific scenario. 
As these empirical filters were just qualitatively adjusted in \cite{Martinez19}, 
the reproduction of their high-pass nature and their order of magnitude is more important than the specific MSE values.}
\vspace{0.2cm}

The \blue{similarity} of the structure of the empirical and theoretical interaction matrices (Eq. \ref{new_kernel_eq} and Eq. \ref{relation_W_H}), and the coincidence of empirical and theoretical filters (Fig. \ref{Fig_filters}) suggests that the proposed theory explains the modifications that had to be introduced in classical unit-norm kernels in Divisive Normalization to explain contrast response.

\subsubsection{Shape adaptation of the kernel depending on the signal}

Once we have shown the global high-pass nature of the vectors $\frac{\vect{k}}{\vect{x}}$ and $\frac{\vect{k}}{\vect{b}} \odot \blue{g_n(\vect{x})}$, lets see in more detail the signal-dependent adaptivity of the kernel.
In order to do so, lets consider the interaction neighborhood of two particular sensors
in the wavelet representation of an illustrative stimulus with easy to understand features. Specifically, the sensors highlighted in red and blue in Fig. \ref{explanation}.

\begin{figure}[b!]
	\centering
    \small
    \setlength{\tabcolsep}{2pt}
    \begin{tabular}{c}
    \hspace{-0.0cm} \includegraphics[width=1.2\textwidth]{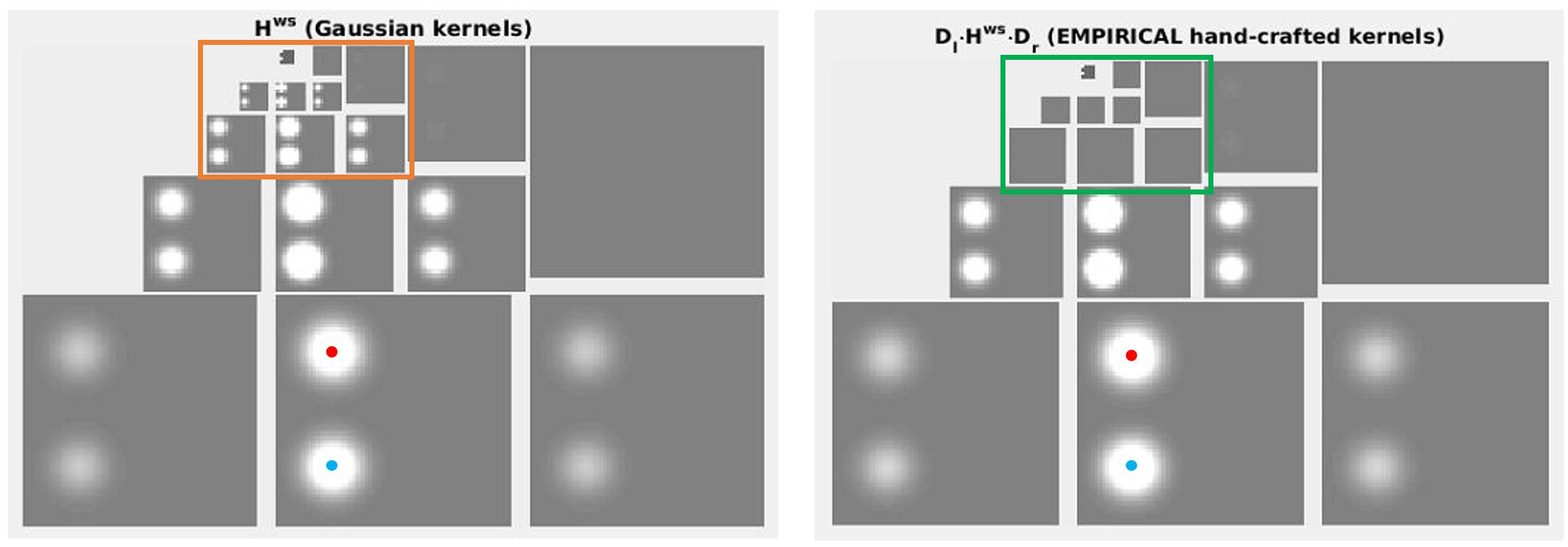} \\
    \end{tabular}
    \vspace{-0.15cm}
	\caption{\textbf{Gaussian and empirical interaction kernels for the sensors highlighted in red and light blue in Fig. \ref{explanation}}. Gaussian kernel (left) with overestimated contribution of low-frequency subbands (highlighted in orange). Hand-crafted kernel (right) to reduce the influence of low-frequencies subbands (highlighted in green).}.
\label{kernels}
    \vspace{-0.15cm}
\end{figure}

Fig. \ref{kernels} compares different versions of the two individual neighborhoods displayed in the same wavelet representation:
\emph{left} the unit-norm Gaussian kernels, $\vect{H}^{ws}$, and \emph{right} the empirical kernel modulated by \emph{ad-hoc} pre- and post-filters, Eq. \ref{new_kernel_eq}.
In these diagrams lighter gray in each $j$-th sensor corresponds to bigger interaction with the considered $i$-th sensor (highlighted in color).
The gray values are normalized to the global maximum in each case.
Each subband displays two Gaussians. Obviously, each Gaussian corresponds to only one of the sensors (the one highlighted in red or in blue, depending on the spatial location of the Gaussian). We used a single wavelet diagram since the two neighborhoods do not overlap and there is no possible confusion between them.

In the base-line unit-norm Gaussian case, $\vect{H}^{ws}$, a unit-volume Gaussian in space is defined centered in the spatial location preferred by the $i$-th sensor. Then, the corresponding Gaussians at every subband are weighted by a factor that decays as a Gaussian over scale and orientation from the maximum, centered at the subband of the $i$-th sensor.

The \emph{problem} with the unit-norm Gaussian in every scale is that the reduced set of sensors for low-frequency scales
lead to higher values of the kernel so that it has the required volume.
In that situation the impact of activity in low-frequency subbands is substantially higher. This fact, combined with the low-pass trend of wavelet signals, implies a strong bias of the response and ruins the contrast masking curves.
This \emph{problem} is represented by the relatively high values of the neighborhoods in the low-frequency subbands highlighted in orange.



This overemphasis in the low-frequency scales was corrected \emph{ad-hoc} using right- and left- multiplication in Eq. \ref{new_kernel_eq} by hand-crafted high-pass filters. The effect of these filters is to reduce the values for the Gaussian neighborhoods at the low-frequency scales, as seen in the empirical kernel at Fig. \ref{kernels}-right. The positive effect of the high-pass filters is reducing the impact of the neighborhoods at low-frequency subbands (highlighted in green).

In both cases (the classical $\vect{H}^{\vect{ws}}$, and the hand-crafted $\vect{H} = \mathbb{D}_{\vect{l}} \cdot \vect{H}^{\vect{ws}} \cdot \mathbb{D}_{\vect{r}}$) the size of the interaction neighborhood (the interaction length) is signal independent. Note that the neighborhoods for both sensors (red and blue) are the same, regardless of the different stimulation that can be seen in Fig. \ref{explanation}.

\begin{figure}[t]
    \vspace{-0.0cm}
	\centering
    \small
    \setlength{\tabcolsep}{2pt}
    \begin{tabular}{c}
    \hspace{-2cm} \includegraphics[width=1.45\textwidth]{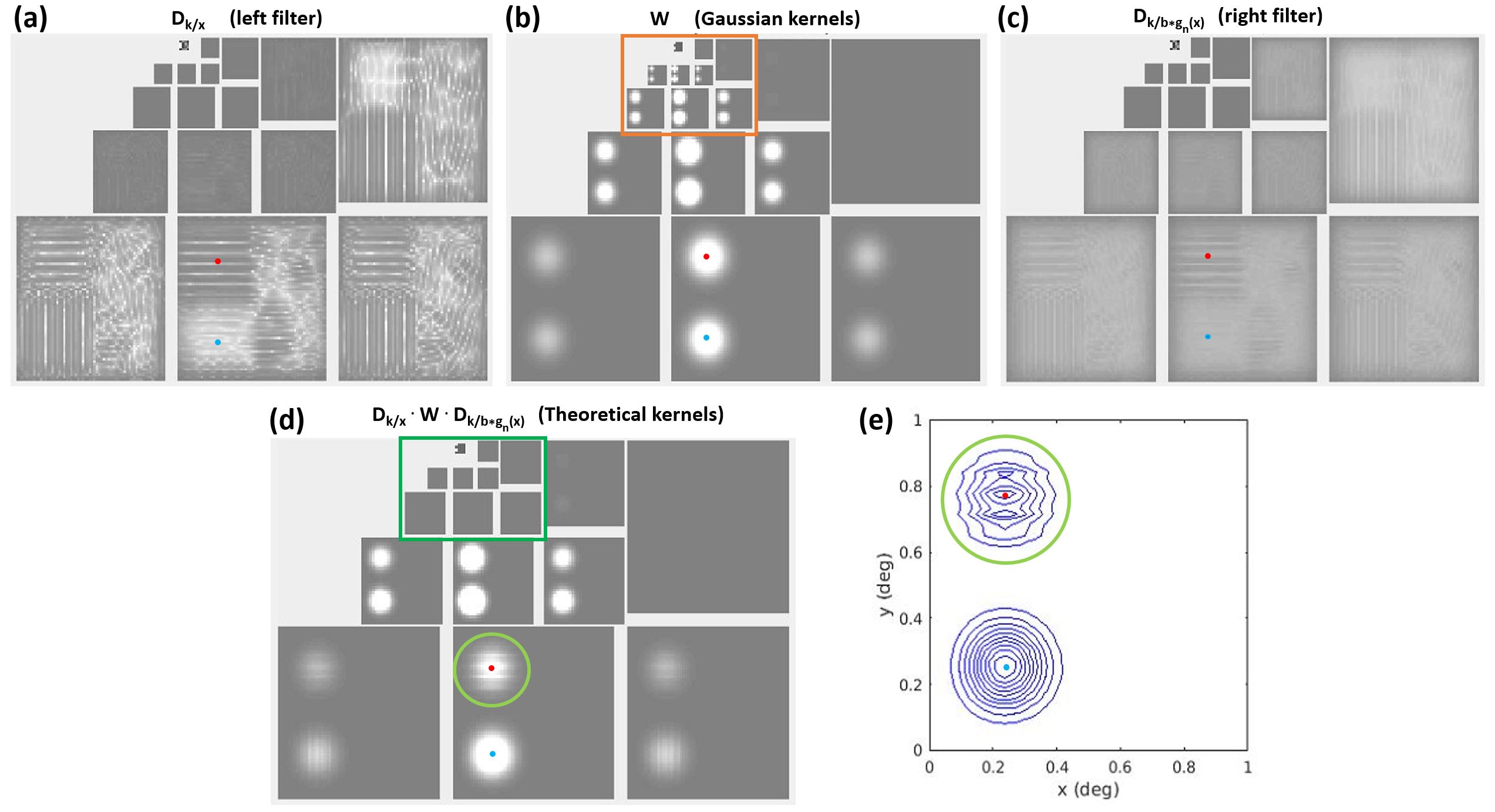} \\
    \end{tabular}
    \vspace{-0.0cm}
	\caption{\small{\textbf{Changes in the shape of the interaction in the theoretically-derived kernel.} Panels \textbf{(a)}, \textbf{(b)}, and \textbf{(c)} show the isolated factors in the kernel matrix $\vect{H}$,
 assuming a Gaussian wiring in $\vect{W}$. The Gaussian component implies interactions with low frequencies (highlighted in orange). \textbf{(d)} Shows the interaction kernel resulting from the product of the three factors  $\vect{H} = \mathbb{D}_{\left(\frac{\vect{k}}{\vect{x}}\right)} \cdot \vect{W} \cdot \mathbb{D}_{\left(\frac{\vect{k}}{\vect{b}} \odot g_n(\vect{x})\right)}$, for the two highlighted points. Here we used $g_{10}(\vect{x})$. Note the high-pass effect of the left- and right- matrix product over $\vect{W}$, that removed the interaction in the low-frequency subbands, now highlighted in dark green. \textbf{(e)} Zoom on the high-frequency horizontal subband. The term depending on the derivatives implies changes of the shape of the kernel (from circular to horizontal ellipses) when the context is a high contrast horizontal pattern. This is compatible with the probabilities of co-assignment~\cite{Coen12} recalled in Fig.~\ref{fig_cohen}.b.}}\label{lo_nuestro}
    \vspace{-0.15cm}
\end{figure}

Fig.~\ref{lo_nuestro} shows the kernels obtained from Eq.~\ref{relation_W_H}.
The three components of $\vect{H}$ are: in Fig.~\ref{lo_nuestro}.a the term proportional to $\frac{1}{\vect{x}}$, in Fig.~\ref{lo_nuestro}.b~the term based on Gaussian neighborhoods $\vect{W}$, and in Fig.~\ref{lo_nuestro}.c~the term proportional to $g_n(\vect{x})$. Finally, Fig.~\ref{lo_nuestro}.d shows the global result of the product of the three terms and the Fig.~\ref{lo_nuestro}.e zooms on the high-frequency horizontal subband that contains the co-linear situation considered in the physiological experiments~\cite{Cavanaugh02b}.

These three terms have the following positive results:
(1)~the product by the high-pass terms moderates the effect of the unit-norm Gaussian at low-frequency subbands as in the empirical kernel tuned in~\cite{Martinez19} shown in Fig.~\ref{kernels}-right, (2)~the term proportional to $\frac{1}{\vect{x}}$ scales the interaction length according to the signal, and (3)~the shape of the kernel depends on the signal because $H_{ij}$ is modulated by $(g_n(\vect{x}))_j$, and this implies that when the surround is aligned with the sensor, the kernel elongates in that direction (as the probability of co-assignment in Fig.~\ref{fig_cohen}.b).
This will lead to smaller responses when the sensor is flanked by co-linear stimuli (as in Cavanaugh et al. results~\cite{Cavanaugh02b}).
\vspace{0.2cm}

In summary, deriving the Divisive Normalization as the steady state of a Wilson-Cowan system with Gaussian unit-norm wiring explains two experimental facts: (1)~the high-pass filters that had to be added to the structure of the kernel in Divisive Normalization to reproduce contrast responses \cite{Martinez19}, and (2)~the adaptive asymmetry of the kernel that changes its shape depending on the background texture~\cite{Nelson1985,Deangelis94, Walker99,Cavanaugh02a,Cavanaugh02b}.

\subsubsection{Contrast response curves from the Wilson-Cowan model}

The above results suggest that the Wilson-Cowan model could successfully reproduce contrast response curves and masking,
which have not yet been addressed through this model. Here we explicitly check this hypothesis.

We can use the proposed relation, Eq. \ref{relation_W_H}, to plug successful parameters of Divisive Normalization tuned
for contrast perception into the \blue{corresponding} Wilson-Cowan model.
We can avoid the integration of the differential equation using the knowledge of the steady state.
The only problem to compute the response through the steady state solution is that the kernel of the Divisive Normalization
depends on the (still unknown) response.

In this case we compute a first guess of the response, $\hat{\vect{x}}$, using the fixed hand-crafted kernel
tuned in \cite{Martinez19}, and then, this first guess is used to compute the proposed signal-dependent kernel, which in turn
is used to compute the actual response, $\vect{x}$.

\begin{figure}[t]
    \small
    \setlength{\tabcolsep}{2pt}
    \begin{tabular}{c}
     \hspace{-1cm} \includegraphics[width=1.2\textwidth]{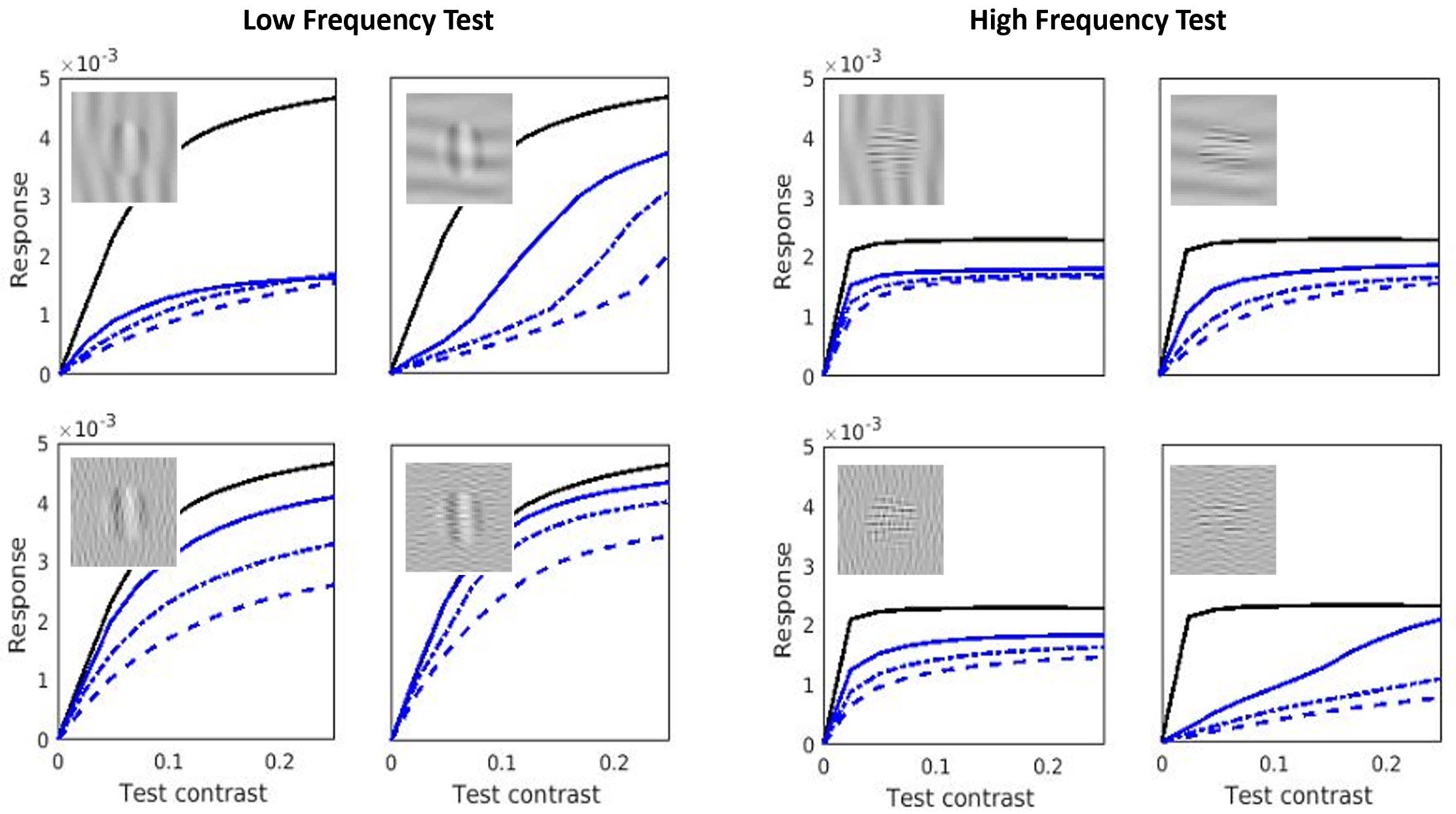} \\
    \end{tabular}
    \vspace{-0.05cm}
	\caption{\small{\textbf{Contrast response curves obtained from the Wilson-Cowan model.} Contrast response curves for low spatial frequency vertical tests (left) and high spatial frequency horizontal tests (right) seen on top of backgrounds of different spatial frequencies, orientations, and contrasts
(see representative stimuli in the insets). The backgrounds include: (1)~two spatial frequencies (low and high, corresponding to the top and bottom rows, respectively); (2)~two orientations (vertical and horizontal, as seen in the insets); and (3)~four different contrasts represented by the line styles (0.0, 0.15, 0.30, and 0.45, corresponding to the black solid line, blue solid line, dotted blue line, and dashed blue line, respectively).
The responses display the qualitative trends of contrast perception: frequency selectivity, saturation with contrast, and cross-masking depending on spatio-frequency similarity between test and background.}}
\label{contrast_curves}
    \vspace{-0.15cm}
\end{figure}

Fig. \ref{contrast_curves} shows the response curves corresponding to neurons that are tuned to low and high spatial frequency tests,
as a function of the contrast of these tests located on top of backgrounds of different contrast, spatial frequency, and orientation.
In each case we considered four different contrasts for the background (represented by the different line styles).
Representative stimuli are shown as image patches inside each plot.
The results in this figure display the expected qualitative properties of contrast perception:

\paragraph*{Frequency selectivity.}  \hspace{0.15cm} The magnitude of the response depends on the frequency of the test: responses for the low-frequency test are bigger than the responses for the high-frequency test. This frequency-dependent behavior in Fig.~ \ref{contrast_curves} is consistent with the Contrast Sensitivity Function~\cite{Campbell68}.

\paragraph*{Saturation.}  \hspace{0.15cm}  The responses increase with the contrast of the test, but this increase is non-linear (saturates), and the responses decrease with the contrast of the background.
This behavior in Fig.~\ref{contrast_curves} is consistent with the contrast discrimination results in~\cite{Legge80,Legge81}.

\paragraph*{Cross-masking.}  \hspace{0.15cm}  Reduction of the responses depends on the frequency similarity between test and background.
Note that the low-frequency test is more attenuated by the low-frequency background of the same orientation than by the high-frequency background of orthogonal orientation. Similarly, the high-frequency test is more affected by the high-frequency background of the same orientation. This behavior in Fig.~\ref{contrast_curves} is consistent with cross-masking results in~\cite{Foley94,Watson97}.

\subsubsection{Metric in the image space from the Wilson-Cowan model}

As a result of the derived relation between models, Eq. \ref{relation_W_H}, the Wilson-Cowan model may also be used to predict subjective image distortion scores.
In this section we explicitly check the performance of the Wilson-Cowan response to compute the visibility of distortions from neural differences following the same approach detailed in the previous section regarding the computation of the signal-dependent kernel and its use to obtain the steady state.

\begin{figure}[t]
	\centering
    \small
    \setlength{\tabcolsep}{2pt}
    \begin{tabular}{c}
     \hspace{-1cm} \includegraphics[width=1.2\textwidth,height=8cm]{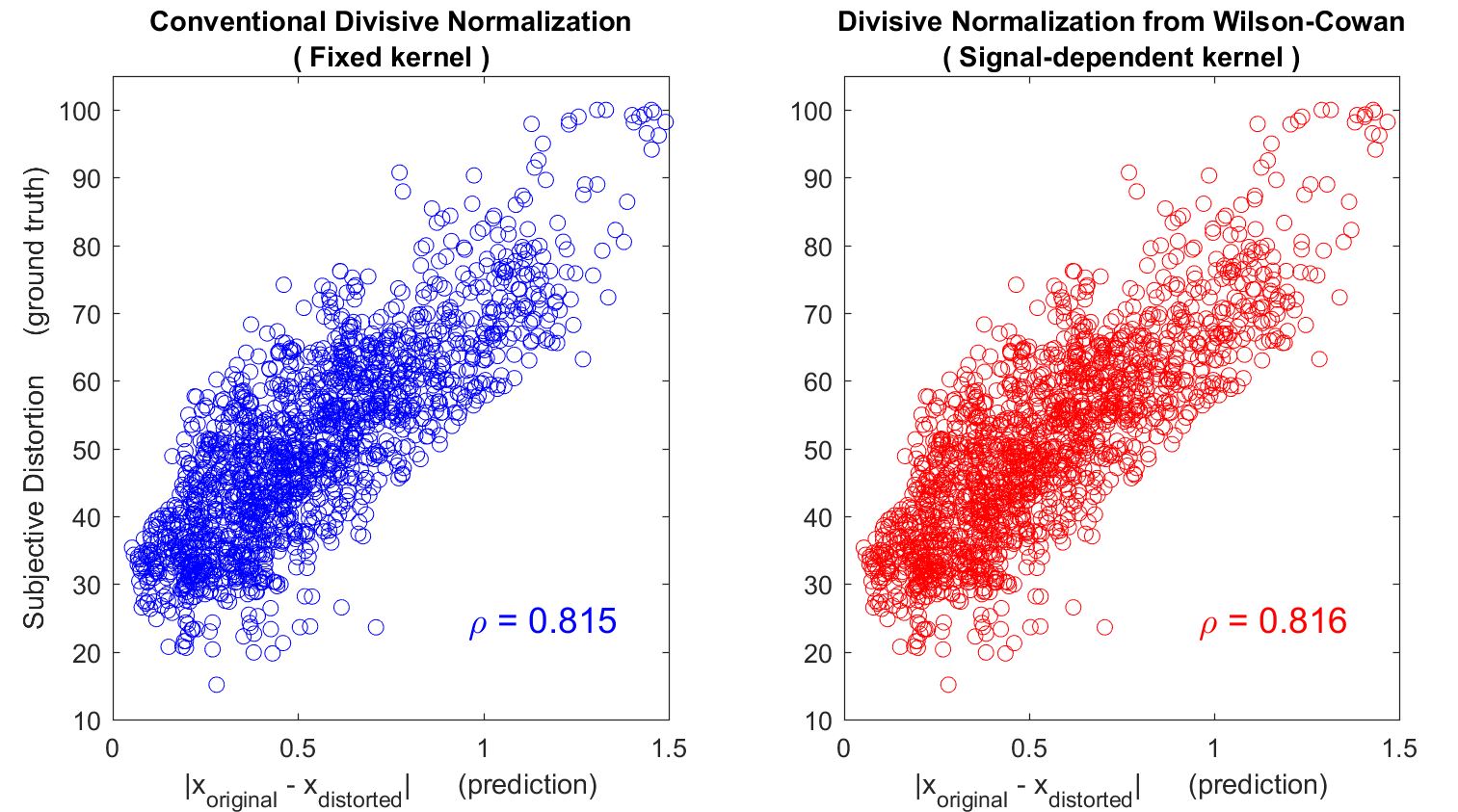} \\
    \end{tabular}
    \vspace{0.05cm}
	\caption{\small{\textbf{Subjective image distortion using the hand-crafted kernel \cite{Martinez19} (left), and the kernel based on Wilson-Cowan equations (right).} For each scatter plot, the Pearson correlation between Mean Opinion Scores (ordinates) and predicted image distortions (abscissas) is given. Differences in the correlations are not statistically significant indicating the \blue{validity} of the proposed relation.}}
	\label{distortion}
    \vspace{-0.15cm}
\end{figure}

The TID database \cite{ponomarenko08,TID2008} contains natural images modified with many kinds of degradation
and has the experimental subjective distortion for each degraded image.
Given a model, the theoretical prediction of the subjective distortion is obtained from the modulus $|\vect{x}_{\textrm{orig}}-\vect{x}_{\textrm{distort}}|$, i.e. the Euclidean difference of the model responses to the original and to the degraded images.

Figure \ref{distortion} compares these predictions (abscissas) with the experimental distortions (ordinates) for the responses with a fixed interaction kernel (the conventional Divisive Normalization approach, in blue), and with the proposed signal-adaptive kernel obtained from the Wilson-Cowan model in red.

The high values obtained for the Pearson's correlation coefficients in both cases, and the close similarities between the plots, prove the good performance of the models and the \blue{validity} of the proposed relation between them.

\section{Final remarks}


In this paper we derived an analytical relation between two well-known models of nonlinear neural interaction: the Wilson-Cowan model~\cite{Wilson72,Wilson73} and the Divisive Normalization model~\cite{Carandini94,Carandini12}.
Specifically, assuming that the Divisive Normalization is the steady state solution of the Wilson-Cowan differential equations,
the Divisive Normalization interaction kernel may be derived from the Wilson-Cowan kernel weighted by two signal-dependent contributions.

We showed the appropriateness of the proposed relation \blue{in a range of model parameterizations} by checking the convergence of the Wilson-Cowan solution to the Divisive Normalization solution,
and by proving that the Divisive Normalization solution is a stable node of the Wilson-Cowan system.

Moreover, the derived relation has the following implications in contrast perception:
(a)~the specific structure obtained for the interaction kernel of Divisive Normalization explains the need of high-pass filters for unit-norm Gaussian interactions to describe contrast masking found in~\cite{Martinez19};
(b)~the signal-dependent kernel predicts elongations of the interaction neighborhood in backgrounds aligned with the sensor,
thus providing a mechanistic explanation to the adaptation facts found in~\cite{Cavanaugh02a,Cavanaugh02b};
and (c)~low-level Wilson-Cowan dynamics may also explain behavioral aspects that have been classically explained through Divisive Normalization,
such as contrast response curves \cite{Foley94,Watson97}, or image distortion metrics \cite{Laparra10a,Berardino17}.
This is the first work that justifies why the Wilson-Cowan interaction successfully reproduces image distortion metrics
and contrast response curves. As stated in~\cite{bertalmio2020visual} there are not many works that explore the use of Wilson-Cowan equations
to model psychophysics, so the examples presented in this work are relevant to fill this gap.

The assumption of the discrete time step $\Delta t$ in the Euler integration of the Wilson-Cowan equations, Eq.~\ref{initial_solution_Euler}, has implications which were not considered in this work. Here the specific value for $\Delta t$ was just an arbitrary choice done for computational convenience. If one could assume that this $\Delta t = 10^{-5}$ is measured in seconds, as the process converges in about 400-500 Euler steps (as seen in Fig. 5), this would mean 4-5 ms to arrive at the steady-state. As an illustration, 4-5 ms would not be a relevant time delay for visual processing of motion, because the cutoff frequency of the temporal Contrast Sensitivity is about 70-100 Hz (so events below 10-15 ms are disregarded), see~\cite{Kelly79}. However, here, the choice for this $\Delta t$ is not based in the biophysics of the neural interactions, and it may indeed be larger~\cite{Zeraati23}. If the interaction time is in fact larger, the convergence will take longer than 4-5 ms. This would imply that the actual behavior would be given by the dynamic Wilson-Cowan model and not by the Divisive Normalization approximation. This may imply that the use of static models like DN should be limited to slow-varying stimuli, or that the use of DN is more correct on a certain region of spatio-temporal frequencies (or speeds). Nevertheless, the detailed analysis of that region from a sensible biophysical estimation of $\Delta t$ is out of the scope here and a matter for further work.

Finally, the \blue{relation} between models proposed here opens the possibility to analyze Divisive Normalization from new perspectives, following methods that have been developed for Wilson-Cowan systems \cite{Sejnowski09}.
Similarly, mechanisms that generalize the Wilson-Cowan equation such as the neurons with intrinsically nonlinear receptive
fields~\cite{Bertalmio20} could be analyzed via information theoretic tools used to quantify the transmission performance of Divisive Normalization~\cite{Serences14,Malo20,Malo22}, or the functional connectivity~\cite{Li_24}.

\vspace{0.3cm}
\noindent \small{\textbf{Acknowledgements:} This work was partially funded by the Spanish Ministerio de Ciencia e Innovaci\'on (MICIIN/FEDER/UE) projects PID2020-118071GB-I00 and PDC2021-121522-C21, and by the Generalitat Valenciana grants GrisoliaP/2019/035 and CIPROM/2021/056.}

\noindent \small{\textbf{Conflict of interest:} The authors declare no conflict of interest.}

\noindent \small{\textbf{Competing Interests:} The authors declare no financial or non-financial interests.}

\noindent \small{\textbf{Author Contributions:} The study was conceived by J.~Malo and discussed with M.~Bertalmío. The simulations were carried out by J.~Malo and J.J.~Esteve-Taboada. The first draft of the manuscript was written by J.~Malo and J.J.~Esteve-Taboada, and all authors commented on posterior versions of the manuscript. All authors read and approved the final manuscript.}

\noindent \small{\textbf{Data Availability:} The datasets generated during and/or analysed during the current study and the code are available in \texttt{http://isp.uv.es/docs/DivNorm\_from\_Wilson\_Cowan.zip}.}
\vspace{-0.3cm}

%

\bibliographystyle{spphys}       
\bibliography{Manuscript}   



\pagenumbering{roman}


\appendix
\section{\blue{Activation functions}}
\label{activations}


\textbf{A.1 Original activation.} \blue{Wilson \& Cowan~\cite{Wilson72,Wilson73} proposed the use of \emph{typical sigmoids} such as the logistic function as appropriate activation functions:}
\begin{equation}
      \blue{f(\vect{x}) = C \left( \frac{1}{1 + e^{-\frac{\vect{x}}{\vect{e}^\star} }} - \frac{1}{2} \right)}
      \label{fx_wc}
\end{equation}
\blue{where $C = \vect{e}^\star \left( \frac{1}{1 + e^{-1}} - \frac{1}{2} \right)^{-1}$, is just a scaling factor of the response and the domain so that $f(\vect{e}^\star) = \vect{e}^\star$.
In the work we selected the constants $\vect{e}^\star$, i.e. we scaled $f(\cdot)$, using the standard deviation of the linear responses of the V1 cells over the subbands. In this way the activation preserves the scale between the different subbands.}

\blue{The derivative of this original activation function, invoked $n$ times in $g_n(\vect{x})$, is given by:}
\begin{equation}
      \blue{\frac{df}{d\vect{x}}(\vect{x}) = \frac{C}{\vect{e}^\star}  
      \frac{e^{-\frac{\vect{x}}{\vect{e}^\star}}}{\left(1+e^{-\frac{\vect{x}}{\vect{e}^\star}}\right)^2}   } 
\end{equation}

\textbf{\blue{A.2 $\gamma$-activation}.} \blue{Following~\cite{Turner2016} that uses power laws to model activation functions, 
we also used another sigmoid inspired in the luminance-brightness nonlinearities of the retinal photoreceptors~\cite{Martinez17}. Due to its similarity to the exponential $\gamma$-correction transforms, it is called $\gamma$-activation or $\gamma$-saturation throughout the work. 
Regular exponential functions were modified around the origin in~\cite{Martinez17} in order to avoid and control the singularity of the derivative in the origin:}
\begin{equation}
    \blue{f(\vect{x})=} 
\begin{cases}
    \blue{\,\,  \mathrm{sign}(\vect{x}) \, C \, |\vect{x}|^\gamma,}  & \blue{\text{if} \,\, |x| \geq \varepsilon} \\
    \,\, \blue{\mathrm{sign}(\vect{x}) \! \left( a \, |\vect{x}| + b \, |\vect{x}|^2 \right),}             & \blue{\text{if} \,\, |x| < \varepsilon} \\
\end{cases}
\end{equation}

\noindent
\blue{where $\gamma < 1$ so that the function saturates, and the constant $C = (\vect{e}^\star)^{1-\gamma}$, ensures that $f(\vect{e}^\star) = \vect{e}^\star$.
In this work $\gamma = 0.6$ and, as stated for the original activation, the scaling $\vect{e}^\star$ was selected to preserve the relation between the standard deviations of the different subbands in the linear responses. The neighborhood, $\varepsilon$, was selected to be close to the origin $\varepsilon = 1\cdot 10^{-3} \vect{e}^\star$. 
The constants, $a$ and $b$ are chosen to ensure continuity of the derivative at $\varepsilon$: $a = (2-\gamma) \, C \, \varepsilon^{\gamma-1}$, and $b = (\gamma-1) \, C \, \varepsilon^{\gamma-2}$.}

\blue{The derivative of this function, useful to compute $g_n(\vect{x})$, is:}

\begin{equation}
    \blue{\frac{df}{d\vect{x}}(\vect{x})=} 
\begin{cases}
    \blue{\,\,  \gamma \, C \, |\vect{x}|^{\gamma-1},}  & \blue{\text{if} \,\, |x| \geq \varepsilon} \\
    \,\, \blue{a + 2 \, b \, |\vect{x}|,}             & \blue{\text{if} \,\, |x| < \varepsilon} \\
\end{cases}
\end{equation}

\blue{Figure~\ref{param} in the main text shows that both functions share the same qualitative properties: saturating sigmoids with derivatives that peak at the origin and decrease with the signal.}

\section{Matlab code}
\label{code}

This appendix lists the main Matlab routines associated to each experiment described in the main text.
All the material is in \texttt{http://isp.uv.es/docs/DivNorm\_from\_Wilson\_Cowan.zip}.
Detailed parameters of the models and the instructions on how to use these routines are given in the corresponding \texttt{*.m} files.

\begin{itemize}

\item \textbf{The Divisive Normalization retina-cortex model:}
The {\small \texttt{Matlab}} toolbox that implements the 4-layer network for spectral or color images considered in Fig.~1
is in {\small \texttt{BioMultiLayer\_L\_NL\_color.zip}}.
This toolbox includes the model, its inverse and Jacobians, and a distortion metric based on the model.

The file {\small \texttt{demo\_deep\_DN\_iso\_color\_spectral.m}} shows how to choose the parameters of the model, how to apply it to
spectral images and images in opponent color representations, and how to compute the responses, the Jacobians and the inverse.
The demo function {\small \texttt{demo\_metric\_deep\_DN\_iso\_color.m}} shows how to represent conventional digital images in the appropriate
opponent color representation.
\vspace{0.2cm}

\item \textbf{Psychophysically-sensible parameters for the Wilson-Cowan model:}
    The toolbox includes the functions \small{\texttt{saturation\_f.m}} and \small{\texttt{inv\_saturation\_f.m}} that compute and
    invert the dimension-wise saturating response of the Wilson-Cowan model depicted in Fig.~\ref{Parameters} of the main text.
    These functions also compute the corresponding derivative with regard to the stimuli.
    The routine \small{\texttt{Converg\_Stability\_ND\_excit\_inhibit.m}} defines and plots the parameters of the Wilson-Cowan model based on psychophysically-tuned Divisive Normalization.
\vspace{0.2cm}

\item \textbf{Experiments on convergence:}
    The connectivity $\mathbf{W}$ and the activation $f$ are applied together in \small{\texttt{integrability\_WC\_with\_f\_after\_KindReview.m}}
    to check the convergence of the system.
    That script applies Euler integration and shows the convergence of the dynamic solution to the corresponding Divisive Normalization solution.
\vspace{0.2cm}

\item \textbf{Experiments on stability:} The stability of the dynamic Wilson-Cowan system is studied through the eigen-decomposition of the Jacobian that controls the amplification of the perturbations in (\small{\texttt{Converg\_Stability\_ND\_excit\_inhibit.m}}), which includes visualizations of the phase diagram.
\vspace{0.2cm}

\item \textbf{Signal-dependent kernel:} The script \small{\texttt{signal\_dependent\_kernel\_with\_f.m}} generates an illustrative image made of high contrast patterns with selected frequencies to stimulate specific subbands of the models. Then, it computes the responses to such stimulus and the corresponding signal dependent-filters according to the relations derived in the main text, Eq.~\ref{relation_W_H}.
    These theoretical filters are compared to the empirical filters found in~\cite{Martinez19}. Finally, in environments where the surround is aligned with the wavelet sensors, the shape of the interaction kernel is found to change as in~\cite{Cavanaugh02b}.
\vspace{0.2cm}

\item \textbf{Contrast response curves:} The script \small{\texttt{contrast\_response\_WC.m}} generates a series of noisy Gabor patterns of controlled frequency and contrast displayed on top of noisy sinusoids of different frequencies, orientations and contrasts. it computes the visibility of these patterns seen on top of the backgrounds by applying the Divisive Normalization model with the signal-dependent lernel derived from the Wilson-Cowan model. The visibility was computed from the response of the neurons tuned to the tests.
\vspace{0.2cm}

\item \textbf{Image distortion metric:} The series of scripts \small{\texttt{images\_TID\_atd\_thorugh\_WC\_model\_x.m}} compute the Divisive Normalization response with the signal-dependent kernel derived from the Wilson-Cowan model for the original and distorted images of the TID database (previously expressed in the appropriate ATD color space). Then, the Euclidean distance is applied to compute the visibility of the distortions.
    The distances are computed by applying \small{\texttt{metric\_deep\_DN\_iso\_colorWC.m}} that computes the responses by calling \small{\texttt{deep\_model\_DN\_iso\_colorWC.m}}.

\end{itemize}

\section{Small-scale Divisive Normalization}
\label{small_DN}

\paragraph*{Overview.} \hspace{0.15cm}  Similarly to~\cite{Gomez19} the code includes a reduced-scale model which can be applied to 3-pixel images for full visualization of the responses and the phase-space. This reduced-scale model consists of two \emph{linear+nonlinear} layers: (1) a linear \emph{radiance-to-luminance} transform using a standard Spectral Sensitivity Function, $V_\lambda$,
in the spectral integration~\cite{Stiles82}, followed by a simple exponential for the \emph{luminance-to-brightness} nonliniearity applied pixel-wise in the spatial domain, that simulates the Weber-Fechner response to luminance~\cite{Fairchild13},
and (2) a \emph{linear+nonlinear} layer in which the linear transform is a discrete cosine transform (a orthonormal rotation) followed by a
low-pass weighting function that simulate frequency-tuned sensors and the Contrast Sensitivity Function (CSF)~\cite{Campbell68}. Then, the outputs of the frequency sensors undergo a nonlinear interaction that may be a Divisive Normalization~\cite{Carandini94,Carandini12,Martinez17,Martinez19}, or its equivalent Wilson-Cowan network, with parameters computed according to Eq.~\ref{relation_W_H}.

\begin{equation}
  \xymatrixcolsep{2pc}
  \xymatrix{ \vect{x}^0 \ar@/^1pc/[r]^{\scalebox{1.00}{$\mathcal{L}^{(1)}$}} & \vect{r}^1  \ar@/^1pc/[r]^{\scalebox{1.00}{$\mathcal{N}^{(1)}$}} & \vect{x}^1 \ar@/^1pc/[r]^{\scalebox{1.00}{$\mathcal{L}^{(2)}$}}  & \vect{r}^2 \ar@/^1pc/[r]^{\scalebox{1.00}{$\mathcal{N}^{(2)}$}} & \vect{x}^2
}
\label{modular}
\end{equation}

\paragraph*{Transform.}  \hspace{0.15cm}  The actual inputs of our code are the responses of the linear photoreceptors: 3-pixel image vectors with normalized luminance values, i.e. $\vect{r}^1 \in \mathbb{R}^3$.
The normalized luminance was computing dividing the absolute luminance in $cd/m^2$ by the value corresponding to the 95\% percentile
of the luminance, in our case 260 $cd/m^2$.

\begin{itemize}

\item The \emph{luminance-to-brightness} transform, $\mathcal{N}^{(1)}$, is just:
\begin{equation}
      \vect{x}^1 = (\vect{r}^1)^\gamma \,\,\,\,\,\,\,\,\,\, \textrm{where} \,\,\,\, \gamma = 0.6
\end{equation}

\item The linear transform of frequency-tuned sensors with CSF gain, $\mathcal{L}^{(2)}$, is:
\begin{equation}
      \vect{r}^2 = G_{\textrm{CSF}} \cdot F \cdot \vect{x}^1 \,\,\,\,\,\,\,\,\,\, \textrm{where}
\end{equation}
\vspace{-0.5cm}
\begin{eqnarray*}
  F &=& \left(
    \begin{array}{ccc}
      \sqrt{\frac{1}{3}}  & \sqrt{\frac{1}{3}} & \sqrt{\frac{1}{3}} \\[0.3cm]
      \sqrt{\frac{1}{2}}  & 0 & - \sqrt{\frac{1}{2}} \\[0.3cm]
      -\sqrt{\frac{1}{6}} & \sqrt{\frac{2}{3}} & -\sqrt{\frac{1}{6}} \\
    \end{array}
  \right)\\[0.3cm]
  G_{\textrm{CSF}} &=& \left(
    \begin{array}{ccc}
     \,\, 1 \,\,\,\, & \,\,\,\, 0 \,\,\,\,& \,\,\,\, 0 \,\, \\
     \,\, 0 \,\,\,\, & \,\,\,\, 0.5 \,\,\,\,& \,\,\,\, 0 \,\, \\
     \,\, 0 \,\,\,\, & \,\,\,\, 0 \,\,\,\,& \,\,\,\, 0.3 \,\,
    \end{array}
  \right)
\end{eqnarray*}

\item The \emph{Divisive Normalization} of the frequency-tuned sensors, $\mathcal{N}^{(2)}_{\textrm{DN}}$, is:
\begin{equation}
    \vect{x}^2 = sign(\vect{x}^2) \odot \mathbb{D}_{\vect{k}} \cdot \mathbb{D}^{-1}_{\left( \vect{b} + \vect{H} \cdot |\vect{r}^2|^\gamma \right)} \cdot |\vect{r}^2|^\gamma  \,\,\,\,\,\, \textrm{where} \,\,\,\,\,\, \gamma = 0.7, \,\,\,\,\,\,  \textrm{and,}
    \label{DN_B2}
\end{equation}
\vspace{-0.5cm}

{\hspace{-1cm}\parbox[b]{1.5\textwidth}{
\begin{eqnarray*}
  \mathbb{D}_{\vect{k}} &=& \left(
    \begin{array}{ccc}
      0.18 \, & \, 0 \,& \, 0  \\
      0 \, & \, 0.03 \,& \, 0  \\
      0 \, & \, 0 \,& \, 0.01
    \end{array}
  \right)\\
  H &=& \mathbb{D}_l \cdot W \cdot \mathbb{D}_r = \left(
    \begin{array}{ccc}
      0.06 \, & \, 0    \,& \, 0  \\   
      0    \, & \, 0.35 \,& \, 0  \\
      0    \, & \, 0    \,& \, 0.27
    \end{array}
  \right)
  \cdot
  \left(
    \begin{array}{ccc}
      0.93 \, & \, 0.06 \,& \, 0.01  \\
      0.04 \, & \, 0.93 \,& \, 0.05  \\
      0    \, & \, 0.02 \,& \, 0.98
    \end{array}
  \right)
  \cdot
  \left(
    \begin{array}{ccc}
      0.95 \, & \, 0    \,& \, 0     \\
      0    \, & \, 0.27 \,& \, 0     \\
      0    \, & \, 0    \,& \, 0.13
    \end{array}
  \right)
\end{eqnarray*}
}
}

and the vector of semisaturations, $\vect{b}$, is:
\begin{equation*}
\vect{b} = \left(
  \begin{array}{c}
    0.08 \\ 
    0.03 \\ 
    0.01 \\ 
  \end{array}
\right)
\end{equation*}

\item The equivalent Wilson-Cowan interaction, $\mathcal{N}^{(2)}_{\textrm{WC}}$, is defined by the differential equation \ref{EqWC2}, where the auto-attenuation, $\vect{\alpha}$, and the interaction matrix, $\vect{W}$, are:
\begin{eqnarray}
\vect{\alpha} &=& \left(
  \begin{array}{c}
    0.41 \\ 
    1.10 \\
    1.30 \\
  \end{array}
\right)\\
  W &=& \left(
    \begin{array}{ccc}
     \, 0.93 \,\,\, & \,\,\, 0.06 \,\,\,& \,\,\, 0.01 \, \\
     \, 0.04 \,\,\, & \,\,\, 0.93 \,\,\,& \,\,\, 0.05 \, \\
     \, 0    \,\,\, & \,\,\, 0.02 \,\,\,& \,\,\, 0.98 \,
    \end{array}
  \right)
\end{eqnarray}
and the saturation function is:
\begin{equation}
      f(\vect{x}) = c \, \vect{x}^\gamma \,\,\,\,\,\, \textrm{where} \,\,\,\,\,\, \gamma = 0.4, \,\,\,\,\,\,  \textrm{and,}
\end{equation}
the scaling constant is, $c = \hat{\vect{x}}^{1-\gamma}$, and $\hat{\vect{x}}$ is the average response over natural images (for the Divisive Normalization transform):
\begin{equation*}
      \hat{\vect{x}} = \left(
  \begin{array}{c}
    1.12 \\ 
    0.02 \\
    0.01 \\
  \end{array}
\right)
\end{equation*}
This exponent is also used for the definition of energy in Wilson-Cowan, $\vect{e} = |\vect{r}|^\gamma$.
\end{itemize}

Note that the interaction neighborhoods have unit volume, $\sum_j W_{ij} = 1 \,\,\, \forall j$, as suggested in~\cite{Watson97}, and then, the Divisive Normalization kernel is given by the product of this unit-volume neighborhood and two left and right filters in the diagonal matrices, $\mathbb{D}_l$ and $\mathbb{D}_r$~\cite{Martinez19}. The values for the semisaturation, $\vect{b}$, and the diagonal matrices $\mathbb{D}_l$ and $\mathbb{D}_r$ were inspired by the contrast response results in~\cite{Martinez19}: we set the semisaturation according to the average response of natural images (low-pass in nature), and we initialized the left and right filters to high-pass.
However, afterwards, in order to make $\mathcal{N}_{\textrm{DN}}$ and $\mathcal{N}_{\textrm{WC}}$ consistent, we
applied the Divisive Normalization over natural images and we iteratively updated the values of the right and left filters
according to Eq.~\ref{relation_W_H}.
In the end, we arrived to the values in the above expressions (where the filter at the left is high-pass, but the filter at
the right is not).
Note that the attenuation in Wilson-Cowan is computed using Eq.~\ref{relation_W_H}.

\paragraph*{Jacobian.}  \hspace{0.15cm}  The perceptual metric \cite{Martinez17} and information transmission \cite{Gomez19,Malo20} strongly depend on how the system (locally) deforms the signal representation. This is described by the Jacobian of the transform with regard to
the signal, $\nabla_{\vect{r}^1} S = \nabla_{\vect{r}^2} \mathcal{N}^{(2)} \cdot \nabla_{\vect{x}^1} \mathcal{L}^{(2)} \cdot \nabla_{\vect{r}^1} \mathcal{N}^{(1)}$. In this reduced-scale model, this Jacobian (for the Wilson-Cowan case) is:
\begin{equation}
      \nabla_{\vect{r}^1} S = \gamma_2 \left(\mathbb{D}_{\vect{\alpha}} + \vect{W} \cdot \mathbb{D}_{\frac{df}{dx}}\right)^{-1}
\cdot \left( \mathbb{D}_{\left( \mathbb{D}_{\vect{\alpha}} \cdot \vect{x}^2 + \vect{W} \cdot f(\vect{x}^2) \right)}\right)^{1-\frac{1}{\gamma_2}} \cdot G_{\textrm{CSF}} \cdot F \cdot \mathbb{D}_{\gamma_1 (\vect{r}^1)^{\gamma_1-1}}
\end{equation}

\end{document}